\documentclass[a4paper,11pt]{article}
\pdfoutput=1
\usepackage{jcappub}

\usepackage[T1]{fontenc}
\usepackage{multirow}
\usepackage{cuted}
\usepackage{amsfonts}
\usepackage{mathtools}
\usepackage{algorithm}
\usepackage{algpseudocode}
\usepackage{lineno}
\usepackage{xcolor}
\usepackage{comment}
\usepackage{dirtytalk}
\usepackage{textcomp}
\usepackage[english]{babel}
\usepackage{aas_macros}

\newcommand{\DifRel}{ \Delta^{\! \mathrm{r}} }
\newcommand{\DifAbs}{ \Delta^{\! \mathrm{a}} }
\newcommand{\SubMax}{ _{\mathrm{max} } }
\newcommand{\dd}{\mathrm{d}}

\title{AutoKnots: Adaptive Knot Allocation for Spline Interpolation}

\author[a]{Sandro D. P. Vitenti,\note{Corresponding author.}}
\author[b]{Fernando de Simoni,}
\author[c,a]{Mariana Penna-Lima,}
\author[d, a]{Eduardo J. Barroso,}

\affiliation[a]{Departamento de F\'isica, Universidade Estadual de Londrina, \\
Rod. Celso Garcia Cid, Km 380, 86057-970, Londrina, Paran\'a, Brazil,}
\affiliation[b]{Departamento de Ci\^encias da Natureza, Universidade Federal Fluminense, \\
Rua Recife, Lotes, 1-7, Jardim Bela Vista, Rio das Ostras - RJ, 28895-532, Brazil,}
\affiliation[c]{Centro Internacional de F\'isica, Instituto de F\'isica, Universidade de Bras\'ilia, \\
70910-900, Bras\'ilia, DF, Brasil,}
\affiliation[d]{LAPP, Universit\'e de Savoie, CNRS/IN2P3, Annecy-Le-Vieux, France}

\emailAdd{vitenti@uel.br}
\emailAdd{fsimoni@id.uff.br}
\emailAdd{pennalima@unb.br}
\emailAdd{barroso@lapp.in2p3.fr}

\abstract{ In astrophysical and cosmological analyses, the increasing quality and volume
	of astronomical data demand efficient and precise computational tools. Interpolation
	methods, particularly spline-based approaches, play a critical role in this context.
	This work introduces a novel adaptive algorithm for automatic knots (AutoKnots)
	allocation in spline interpolation, designed to meet user-defined precision
	requirements. Unlike traditional methods that rely on manually configured knot
	distributions with numerous parameters, the proposed technique automatically
	determines the optimal number and placement of knots based on interpolation error
	criteria. This simplifies configuration, often requiring only a single parameter.
	The algorithm progressively improves the interpolation by adaptively sampling the
	function-to-be-approximated, $f(x)$, in regions where the interpolation error
	exceeds the desired threshold. All function evaluations contribute directly to the
	final approximation, ensuring efficiency. While each resampling step involves
	recomputing the interpolation table, this process is highly optimized and usually
	computationally negligible compared to the cost of evaluating $f(x)$. However, for
	inherently fast functions, interpolation may not offer significant benefits. We show
	the algorithm's efficacy through a series of precision tests on different functions.
	However, the study underscores the necessity for caution when dealing with certain
	function types, notably those featuring plateaus. To address this challenge, a
	heuristic enhancement is incorporated, improving accuracy in flat regions.
	Originally developed in 2007 and integrated into the Numerical Cosmology library
	(NumCosmo), this algorithm has been extensively used and tested over the years.
	NumCosmo includes a comprehensive set of unit tests that rigorously evaluate the
	algorithm both directly and indirectly, underscoring its robustness and reliability.
	As a practical application, we compute the surface mass density $\Sigma(R)$ and the
	average surface mass density $\overline{\Sigma}(<R)$ for Navarro-Frenk-White and
	Hernquist halo density profiles, which provide analytical benchmarks. The adaptive,
	slim algorithm is implemented in NumCosmo, offering a mature and user-friendly tool
	for interpolation challenges in computational astrophysics and cosmology. }

\begin{document}
\maketitle
\flushbottom

\section{Introduction}

Cosmology and astrophysics are entering a high-precision era driven by the large influx
of high-quality data from large-scale surveys. These ongoing and near-future galaxy
surveys, such as the Dark Energy Spectroscopic Instrument (DESI) \cite{Levi2019}, the
Javalambre Physics of the Accelerating Universe Astrophysical Survey
(J-PAS)~\cite{Benitez2014}, Euclid~\cite{Euclid2024}, the Rubin Observatory Legacy
Survey of Space and Time (LSST)~\cite{LSST2012} and the Nancy Grace Roman
Telescope~\cite{Dore2019}, encompass a broad range of observational data enabling the
exploration of fundamental questions such as the nature of dark energy and dark matter.

However, as the precision of cosmological and astrophysical analyses improves, the
nature of the challenges shifts. Statistical errors, once a primary concern, are now
becoming less significant than systematic ones. Furthermore, advancements in our
understanding of various dynamic astrophysical and cosmological phenomena make their
numerical modeling increasingly detailed and complex. The Hubble tension exemplifies
this current landscape, as it has been the focus of extensive investigation, involving
the exploration of modeling approaches, systematic errors, and even the consideration of
new physics~\citep{DiValentino2021, Kamionkowski2023, Riess2024}.

A third, but no less important, source of error is the numerical error. The relevance of
well-tested and optimized computational codes cannot be overstated. The complexity of
modern cosmological and astrophysical models demands that the tools used for analysis be
rigorously validated to ensure their accuracy. Any flaws in these codes can introduce
numerical errors, which, if left unchecked, could lead to spurious results that
undermine the integrity of the research. Therefore, the development and maintenance of
reliable software are critical to advancing our understanding of the universe. An
efficient practice to overcome this problem is the distribution of independent and
open-source programming libraries. We can mention some good examples such as
CAMB~\cite{CAMB2011}, CLASS~\cite{CLASS2011}, Colossus~\cite{Colossus2018},
CCL~\cite{CCL2019}, and NumCosmo~\cite{NumCosmo2014}.

As previously noted, the large volume of data and the complexity of modeling
cosmological probes necessitate optimized algorithms to enable feasible statistical
analyses. Optimization methods often rely on techniques to approximate or estimate
target functions and related operations, such as differentiation and integration.
Interpolation techniques, including polynomial, spline, radial basis function
interpolations~\cite{Powell1981, Obradovic2021}, and Gaussian
processes~\cite{Candela2007, Quadrianto2010}, are widely used for this purpose. In
particular, the numerical analyses performed by these libraries often require evaluating
computationally intensive functions. To reduce this workload, interpolation methods are
employed: the expensive function is computed at a limited number of points, while the
remaining values are estimated through interpolation. This approach plays a crucial role
in the efficiency and development of these libraries.

A key consideration when using an interpolation method is determining the number and
placement of knots. Minimizing the number of knots is crucial to reducing the
computational cost of both evaluating the interpolation and the underlying high-cost
function. At the same time, enough knots must be allocated to achieve the desired
precision of the approximation. The placement of knots also affects the quality of the
interpolation, as poorly chosen positions can lead to inefficiencies or inaccuracies in
regions where the function changes rapidly~\cite{lyche1999sensitivity}. Various
techniques for optimizing knot allocation have been studied~\cite{michel2021new,
	goepp2018spline, li2005adaptive}. For example, \cite{goldenthal2004spline} propose
using a genetic algorithm to minimize a knot-dependent cost function,
\cite{galvez2015elitist} apply an elitist clonal selection algorithm, and
\cite{idais2019optimal} utilize a Multi-Objective Genetic Algorithm. These approaches
aim to balance computational efficiency with the precision of the interpolated function.

In most cases, users must manually configure the knot distribution, which can be both
challenging and time-consuming. This process involves selecting the number and positions
of knots, a task that becomes particularly difficult when the function to be
interpolated is unknown or exhibits complex behavior. In cosmology and astrophysics, the
optimal knot distribution often depends on the underlying model. For instance, the
cosmological distance-redshift relation varies with cosmological parameters, and the
number of knots needed for interpolation can change accordingly.

This dependence complicates the determination of an optimal knot distribution,
especially during statistical analyses like Markov Chain Monte Carlo (MCMC), where model
parameters are sampled from a probability distribution. In such cases, the knot
configuration must ensure the desired precision across the full range of sampled
parameter values. Moreover, in practical analyses in these fields, tens of functions
often require interpolation, leading to a proliferation of configuration parameters that
further increase the complexity and workload for the user.

In this work, we present a deterministic adaptive knot allocation method guided by two
convergence criteria and a specified interpolation error threshold. This method ensures
that in statistical analyses, such as Markov Chain Monte Carlo, the approximated
function achieves the desired precision consistently across the entire parameter space.
The algorithm is implemented in the Numerical Cosmology library
(NumCosmo~\cite{NumCosmo2014}), an open-source library dedicated to cosmological and
astrophysical computations, as well as statistical tools for data
analysis.\footnote{\url{https://github.com/NumCosmo/NumCosmo}} Written in C and equipped
with a Python interface, the library is designed to be accessible to a wide range of
users. The algorithm undergoes rigorous testing and validation through extensive unit
tests, ensuring its accuracy and reliability.

This paper is organized as follows. In section~\ref{sec:method}, we describe our
deterministic method for automatic knot placement (AutoKnots), detailing the convergence
criteria, parameters, and available options. In section~\ref{sec:tests}, we evaluate the
algorithm through a series of tests on diverse target functions, highlighting its
performance across different scenarios. In section~\ref{sec:cosmology}, we demonstrate the
method's application in a cosmological context by computing the surface mass density and
mean surface mass density for two halo density profiles. Finally,
section~\ref{sec:conclusion} presents our concluding remarks, while
Appendix~\ref{app:cubicspline} defines cubic splines and outlines the
steps for implementing the algorithm.

\section{The adaptive method}
\label{sec:method}

The AutoKnots selection technique presented in this study is broadly applicable to
various interpolation methods. To demonstrate its effectiveness, we focus on cubic
spline interpolation with the not-a-knot condition~\citep{Boor85}. For further details
on cubic spline interpolation and the algorithm design, see Appendix~\ref{app:cubicspline}.

\subsection{Definitions}

Let $f(x)$ represent the function to be interpolated over the interval $[a, b]$. The set
of strictly increasing knots and their corresponding function values denoted by
$\mathbb{K}^t$:
\begin{equation}
	\mathbb{K}^t = \left\{\left(x_0, f(x_0)\right), \left(x_1, f(x_1)\right), \ldots, \left(x_n, f(x_n)\right)\right\},
\end{equation}
where $x_0 = a$ and $x_n = b$, $n+1$ is the number of knots, and $t$ is the iteration
number. Using these, the approximated function for a given interpolation technique is
expressed as:
\begin{equation}
	\hat{f}^t(x) \equiv \hat{f}\left(x|\mathbb{K}^t\right).
\end{equation}

The interval $[a, b]$ is divided into $n$ sub-intervals based on the knots, with their
sizes defined as:
\begin{equation}
	\mathbb{H}^t = \{h_0, h_1,\dots, h_{n-1}\}, \qquad h_i \equiv x_{i+1} - x_i.
\end{equation}
To test the accuracy of the interpolation $\hat{f}^t(x)$, we compare it to the true
function $f(x)$ by evaluating both at a new set of points. To select
these points, we use the midpoint rule, that is, for a given sub-interval $[x_i, x_{i+1}]$,
we define the midpoint $\overline{x}_i$ as:
\begin{equation}
	\overline{x}_i = \frac{x_i + x_{i+1}}{2}.
\end{equation}
The accuracy of the interpolation is then measured by the absolute error between the
true function and its approximation:
\begin{equation}\label{eq:abs_error}
	\DifAbs f^t(x) \equiv \left\vert f(x) - \hat{f}^t_n(x) \right\vert.
\end{equation}
For cubic spline interpolation with the not-a-knot condition, the error typically scales
as $\mathcal{O}(h^4)$, where $h$ is the size of the sub-interval. However, the exact
error also depends on the specific function being interpolated and the sub-interval,
with a proportionality constant that varies accordingly.

An additional test involves comparing the integral of the true function over the
sub-intervals with the integral of the interpolated function. Although we do not have
the exact integral, we can approximate it using Simpson’s 1/3 rule (see, for example,
\citep{davis2014}):
\begin{equation}\label{eq:simpson_rule}
	\widetilde{\mathcal{I}}_i = \frac{h_i}{6} \left[ f(x_i) + 4 f(\overline{x}_i) +
		f(x_{i+1}) \right].
\end{equation}
This approximation also has an error of $\mathcal{O}(h^4)$, allowing us to compare the
interpolation error with the integral error. The key advantage of this approach is that
it provides a different proportionality constant, which helps identify regions where the
interpolation is less accurate than the integral approximation or vice versa. We represent
the integral of the interpolated function and its absolute difference from the Simpson’s
1/3 rule approximation as
\begin{equation}
	\hat{\mathcal{I}}^t_i \equiv \int_{x_i}^{x_{i+1}} \hat{f}^t(x)\,\dd x, \qquad
	\DifAbs \mathcal{I}^t_i \equiv \left| \widetilde{\mathcal{I}}_i - \hat{\mathcal{I}}^t_i \right|.
\end{equation}

Given a tolerance $\delta$ and a scale parameter $\varepsilon$, we define a point as
having an interpolation error within the required precision if the following conditions
are satisfied:
\begin{subequations}
	\label{eq:convergence}
	\begin{align}
		\label{eq:cond1}
		\DifAbs f^t(\overline{x}_i) & < \delta \left[ |f(\overline{x}_i)| + \varepsilon \right],                            \\
		\label{eq:cond2}
		\DifAbs \mathcal{I}^t_i     & < \delta \left[ \left|\widetilde{\mathcal{I}}_i \right| + \varepsilon \, h_i \right].
	\end{align}
\end{subequations}

We track the status of each sub-interval using a set of integer variables
\begin{equation}
	\mathbb{S}^t = \{s^t_0, s^t_1, \ldots, s^t_{n-1}\},
\end{equation}
where $s_i$ represents whether the sub-interval $[x_i, x_{i+1})$ is well approximated by
the interpolation using the current set of knots. If $s^t_i = s_\mathrm{conv}$, the
sub-interval is considered well approximated; if $s^t_i < s_\mathrm{conv}$, the
sub-interval requires further subdivision.

\subsection{Adaptive Algorithm}

The AutoKnots algorithm is designed to automatically determine the optimal number and
placement of knots based on the interpolation error criteria defined in
eqs.~\eqref{eq:convergence}. Under the not-a-knot condition, our cubic spline algorithm
requires a minimum of six knots for interpolation~\citep{Boor85}. In the initial step,
we distribute these six knots uniformly across the interval $[a, b]$, forming the
initial set $\mathbb{K}^0$.

Since the behavior of the function $f(x)$ is unknown -- specifically whether it exhibits
regions of high variability or sharp changes—we begin with a uniform knot distribution.
At this stage, it is uncertain whether $f(x)$ can be accurately approximated by a cubic
spline with six equally spaced knots. To initiate the adaptive process, all $s^0_i$
values are set to zero.

Now, we begin the adaptive process. Given the sets $\mathbb{K}^{t}$ and
$\mathbb{S}^{t}$, we iterate over the sub-intervals. If $ s_i < s_\mathrm{conv}$, we
apply the midpoint rule to generate a test point $\overline{x}_i$ within the
sub-interval $[x_i, x_{i+1})$. We then update the knot set to $\mathbb{K}^{t+1} =
	\mathbb{K}^{t} \cup \{(\overline{x}_i, f(\overline{x}_i))\}$ and evaluate the error
criteria in eqs.~\eqref{eq:convergence} for both the function and its integral.

After updating to $\mathbb{K}^{t+1}$, the sub-interval $[x_i, x_{i+1})$ is subdivided
into two new sub-intervals:
$$
	[x_j = x_i, x_{j+1} = \overline{x}_i), \qquad [x_{j+1} = \overline{x}_i, x_{j+2} = x_{i+1}).
$$
The set $\mathbb{S}^{t+1}$ is updated as follows: if the conditions in
eqs.~\eqref{eq:convergence} are satisfied for both the function and its integral, the
new sub-intervals are assigned $s_i^{t} + 1$. Otherwise, the value of $s_i^{t}$ is
retained. This can be expressed as:
$$
	s_j^{t+1} = s_{j+1}^{t+1} =
	\begin{cases}
		s_i^{t} + 1, & \text{if conditions in eqs.~\eqref{eq:convergence} are satisfied}, \\
		s_i^{t},     & \text{otherwise}.
	\end{cases}
$$
This process is repeated until all sub-intervals are well approximated, meaning $s_i =
	s_\mathrm{conv}$ for all $i$. The final set of knots $\mathbb{K}^{t+1}$ is then used
to compute the final approximation $\hat{f}^{t+1}(x)$. In the next sections we refer to
the final approximation as $\hat{f}(x)$.

In summary, the algorithm improves the interpolation by adaptively sampling the function
$f(x)$ in subregions where the criteria in eqs.~\eqref{eq:convergence} are not met
within $s_\mathrm{conv}$ iterations. In these intervals, the midpoint rule is used to
generate test points, though other strategies were also explored, such as using
Chebyshev points~\citep{Trefethen2013} or identifying regions of high curvature. Among
these, the midpoint rule proved to be the most efficient and effective approach. The
adaptive process continues until all subregions satisfy the criteria, at which point the
final approximation is computed. Furthermore, we observed that setting $s_\mathrm{conv}
	> 1$ typically results in unnecessarily high accuracy, leading to more knots than
required.

In the algorithm, the two precision parameters are initialized by the user (see the
description of the code in section~\ref{app:ncm_spline_func}). We tested various values for
$s_\mathrm{conv}$ and found that the algorithm is robust when $s_\mathrm{conv} = 1$.
Again, using larger values typically results in more knots than necessary. Therefore,
from this point onward, we will consistently use $s_\mathrm{conv} = 1$.

\subsection{Tolerance and Scale Parameters}
\label{sec:anal_conv_crit}

The inclusion of the scale parameter $ \varepsilon $ in the convergence criteria serves
a well-known purpose in numerical computations: it prevents issues that arise when the
relative error becomes ill-defined. Specifically, if $ \varepsilon = 0 $, the convergence
criteria rely solely on relative differences, that is,
\begin{equation} \label{eq:scale_0_1}
	\DifAbs f (\overline{x}_i) < \delta |f(\overline{x}_i)|, \qquad
	\DifAbs \mathcal{I}_i < \delta \left|\widetilde{\mathcal{I}}_i \right|.
\end{equation}
In regions where $ |f(\overline{x}_i)| \to 0 $, the relative error is undefined, and the
algorithm would continue refining the sub-intervals indefinitely, adding unnecessary
knots without ever satisfying the conditions. This results in over-sampling the region
and inefficient computations.

By introducing $ \varepsilon > 0 $, the convergence criteria gain an absolute threshold
that stabilizes the process when $ |f(\overline{x}_i)| $ is small. In these regions, the
criteria are no longer driven to zero but instead bounded by the user-defined $ \varepsilon $,
avoiding numerical instability and ensuring practical convergence.

For regions where $ |f(\overline{x}_i)| \gg \varepsilon $, the $ \varepsilon $-dependent
terms become negligible, and the criteria revert to being dominated by relative differences.
This ensures that the algorithm maintains appropriate relative accuracy for larger function
values, while remaining robust near zero.

In practice, $ \varepsilon $ allows users to specify the scale of function values that are
relevant to their problem. This simple adjustment is especially useful for functions that
span multiple orders of magnitude, ensuring balanced performance across all regions of the
domain.

An important property of the cubic spline interpolation algorithm (under the not-a-knot
condition) is its behavior under scalar multiplication of the function. If the original
function is scaled by a constant factor, $ f_A(x) = A f(x) $, the interpolated function
also scales proportionally, $ \hat{f}_A(x) = A \hat{f}(x) $. This is a direct result of
the algorithm solving a linear system to determine the cubic polynomials within each
interval. Consequently, the interpolation and its \textit{relative} errors remain invariant
under scalar transformations.

Condition~\eqref{eq:scale_0_1} reflects this invariance: for $f_A(x)$, it becomes
$$
	\DifAbs f_A(x) = |A|\DifAbs f(x) < \delta |A| |f(x)|.
$$
While the absolute difference threshold is scaled by $ |A| $, the relative error and the
knot distribution remain unchanged. This property also ensures that the adaptive method
does not distinguish between $f(x)$ and $-f(x)$, as only the magnitude of the function
influences the placement of knots. Note that the same behavior is observed for the
integral error criterion. When incorporating the scale parameter $\varepsilon$,
condition~\eqref{eq:cond1} transforms to
$$
	\DifAbs f_A(x) = |A|\DifAbs f(x) < \delta |A| (|f(x)| + \varepsilon / |A|).
$$
To preserve the same knot distribution for $ f_A(x) $ and $ f(x) $, $ \varepsilon $ must
scale by $|A|$, that is, $\varepsilon \to |A| \varepsilon$.

When $ |f(\overline{x}_i)| \ll \varepsilon $, the convergence criteria reduce to:
$$
	\DifAbs f (\overline{x}_i) < \delta \varepsilon \, ,
$$
and
$$
	\DifAbs \mathcal{I}_i < \delta \varepsilon \, h_i \, .
$$

In this regime, the error bounds become independent of the actual value of
$f(\overline{x}_i)$. The absolute difference of the function is only required to be
smaller than the product of the relative tolerance, $\delta$, and the scale parameter,
$\varepsilon$. Similarly, the integral error bound corresponds to approximating the
integral as a rectangle with width $h_i$ and height $\varepsilon$. This behavior ensures
that in regions where the function magnitude is much smaller than $\varepsilon$, any
interpolated value is effectively accepted as long as it satisfies these constant
bounds.

This scenario often arises when interpolating functions like the Gaussian distribution,
where tails extend far from the peak. In such regions, the function values are
negligible, and relative error becomes irrelevant. The scale parameter $\varepsilon$
effectively defines an absolute threshold, below which differences are ignored,
prioritizing the efficiency of the interpolation process over precision.

When $\varepsilon$ acts as an absolute tolerance, scaled by the relative tolerance
$\delta$, it sets the baseline for the acceptable error. In the extreme case where
$\varepsilon \to \infty$, the convergence criteria are trivially satisfied at the
initial iteration. This leads to a uniform distribution of knots (11 in total: six
initial and five added in the first iteration) and results in an interpolation that does
not adapt to the function's structure. Consequently, the resulting approximation becomes
suboptimal, ignoring essential variations in the function.

\subsection{Error Analysis}

The error in cubic spline interpolation within a interval of width $h_i$ is bounded by
\citep{Boor78}:
\begin{equation}
	\label{eq:error_cond1}
	\DifAbs f_{i} \sim \frac{1}{384} \left| f^{(4)} \right|_{i} h_{i}^{4} \, ,
\end{equation}
where $ f^{(4)} $ represents the maximum fourth derivative of the function within the
interval. This expression highlights how the interpolation error depends on the fourth
derivative and the interval width $h_i$, with the error scaling as $h_i^4$. This
relationship ensures that reducing $h_i$ significantly decreases the interpolation
error, especially in regions where $f^{(4)}$ is large. Errors tend to be more pronounced
near the endpoints due to the additional constraints imposed by cubic spline
interpolation \citep{SUN2023115039, BEHFOROOZ20068}.

To assess the convergence of the integral condition~\eqref{eq:cond2}, note that it
compares two approximations of the function integral. Simpson's 1/3 rule has an
associated error bound given by \citep{davis2014}:
$$
	\int_{x_i}^{x_{i+1}}\! f(x)\, \mathrm{d}x - \widetilde{\mathcal{I}}_i \sim - \frac{1}{2880} \left| f^{(4)} \right|_{i} h_{i}^{5} \, .
$$
The cubic spline integral error bound within the interval $ h_{i} $ is given by
\citep{Phythian86}:
$$
	\int_{x_i}^{x_{i+1}}\! f(x)\, \mathrm{d}x - \hat{\mathcal{I}}_{i} \sim - \frac{1}{180} \left| f^{(4)} \right|_{i} h_{i}^{5} \, .
$$
This relation also underestimates the error of the cubic spline integral in intervals
near the boundaries, similar to the function approximation. Combining both estimates,
the theoretical absolute difference between these two estimators, \textit{i.e.}, the
error bound of condition~\eqref{eq:cond2}, is given by,
\begin{equation}\label{eq:error_cond2}
	\DifAbs \mathcal{I}_i \sim \frac{1}{192} \left|f^{(4)}\right|_{i} h_{i}^{5}.
\end{equation}

For the entire interval $[a, b]$, the maximum interpolation error takes a similar form
\citep{HALL1968209}:
$$
	\DifAbs f(x)_{[a,b]} \sim \frac{5}{384} \left| f^{(4)} \right|_{[a,b]} \max_{i} \{ h_i \}^{4}.
$$
This expression provides an approximate upper bound for the global interpolation error.
If a uniform knot distribution is assumed, it can be used to estimate the number of
knots required to achieve a target error. For instance, for the function $ f(x) =
	\cos(x) $ over $ 0 \leq x \leq \pi $, the maximum fourth derivative is $ |f^{(4)}| = 1
$. Targeting a maximum interpolation error of $ 10^{-8} $, the optimal interval width
can be approximated as:
\begin{equation}
	\label{eq:error_estimate_knots}
	h \sim \left[ \frac{384}{5} \frac{\DifAbs f(x)_{[0, \pi]}}{|f^{(4)}|_{[0, \pi]}} \right]^{1/4}
	\approx 0.03,
\end{equation}
implying the use of approximately 100 knots.

While this calculation offers insights into the relationship between $ h $, $ f^{(4)} $,
and the error, it presents two significant challenges. First, it requires prior
knowledge of the fourth derivative, which may not be readily available for many
functions. Second, assuming a uniform distribution of knots is often suboptimal,
especially for functions with regions of rapid variation or sharp features.

To address these limitations, the adaptive method applies the error
conditions~\eqref{eq:cond1} and \eqref{eq:cond2} independently to each interval $ h_i $,
relying only on evaluations of the function and its integral. By iteratively adjusting
the partitions $ h_i $ based on local error estimates, the adaptive method naturally
produces a nonuniform distribution of knots, refining them in regions of high
variability and spacing them farther apart in smoother regions. This ensures a more
balanced and accurate approximation compared to a uniform distribution.

Moreover, the error bound for numerical integration combines the interpolation error and
the error from Simpson's rule, both of which scale as $ f^{(4)} \times h_i^4 $ but are
evaluated at different points within the interval. This dual consideration ensures that
the adaptive method efficiently captures the range of errors, minimizing discrepancies
across the interval and enhancing the robustness of both interpolation and integration.

Concerning the convergence condition given in eq.~\eqref{eq:cond2}, Simpson's 1/3 rule
is chosen due to its balance between accuracy and computational efficiency. It is exact
for polynomials of degree two and, as discussed, its error aligns well with that of cubic
spline interpolation.
This ensures a good level of accuracy for the integral estimates.
Additionally, Simpson's rule is straightforward to implement and computationally
efficient, making it suitable for applications where multiple integrations are required.
Therefore, the practical balance between simplicity and accuracy provided by Simpson's
1/3 rule makes it a reliable choice.

Alternative numerical integration rules for condition~\eqref{eq:cond2} would affect
convergence behavior differently. Methods that are less aligned with the cubic spline
integral approximation, such as the trapezoidal or Boole's rules \citep{davis2014},
would result in slower convergence, requiring more iterations than with the proposed
condition, resulting in a less accurate approximation. Conversely, using a numerical
integration method more closely aligned with the cubic spline, such as the modified
Simpson's rule \citep{Ujevic2003, BURG2012}, could lead to faster convergence, requiring
fewer iterations, but with a trade-off of reduced precision in the approximation and
additional computational overhead.

\subsection{The \texttt{refine} option}
\label{sec:refine}

One of the main challenges in automatic knot allocation arises when the function
contains regions with sharp features embedded in an otherwise relatively smooth
background. In such cases, the sharp feature may fall entirely within one sub-interval,
while the function appears nearly flat elsewhere. This issue became evident during
tests, which revealed instances where the method failed to achieve the desired accuracy,
particularly in regions of nearly zero curvature, such as inflection points. Here, the
adaptive method can converge prematurely, effectively overlooking the sharp feature and
producing a suboptimal knot distribution.

A similar issue is encountered in adaptive quadrature methods, where algorithms may fail
to adequately sample regions containing singularities or abrupt changes in the
function's behavior. These failures occur because the algorithm's convergence criteria
can be satisfied without sufficiently refining the partition in these problematic areas,
leading to inaccurate results. Addressing this challenge requires additional safeguards
or modifications to the adaptive method to ensure that regions with sharp features
receive the necessary refinement.

To address this issue, we introduce the \texttt{refine} option, a heuristic approach
designed to improve accuracy in regions where premature convergence might have occurred.
The process begins by running the adaptive method until all sub-intervals satisfy the
convergence criteria in eqs.~\eqref{eq:convergence}, resulting in $s_i =
	s_\mathrm{conv}$ for all sub-intervals.

If the \texttt{refine} option is enabled, we then compute the mean
$\text{\texttt{mean}}(H^{t+1})$ and standard deviation $\text{\texttt{std}}(H^{t+1})$ of
the sub-interval widths, $h_i$, across all sub-intervals. Next, we evaluate the
following condition:
\begin{equation} \label{eq:refine}
	h_i > \text{\texttt{mean}}(H) + \texttt{refine\_ns} \times \text{\texttt{std}}(H),
\end{equation}
where \texttt{refine\_ns} is a user-defined parameter controlling the sensitivity of the
refinement. If this condition is met for a sub-interval $[x_i, x_{i+1})$, we reset $s_i
	= 0$ for that sub-interval and restart the adaptive method.

This refinement procedure is repeated for the number of iterations specified by the
user. By iteratively refining intervals with unusually large widths, the \texttt{refine}
option ensures that regions with sharp features are better resolved, reducing the
likelihood of premature convergence and improving the overall accuracy of the method.

The use of standard deviation is intentional, as its sensitivity to outliers helps
identify intervals with significant deviations from the average width. The user-defined
parameter, \texttt{refine\_ns}, determines the threshold in units of
$\text{\texttt{std}}(H)$ for targeting regions that require additional knots. As
\texttt{refine\_ns} increases, the \texttt{refine} option becomes less sensitive,
affecting only the most extreme partitions. Beyond a certain value, the threshold
effectively disables refinement. Selecting an appropriate \texttt{refine\_ns} requires
balancing sensitivity to outliers with the need to avoid excessive refinement, with
moderate initial values often providing satisfactory results.

The user can specify the number of iterations for grid refinement using the parameter
\texttt{refine}, an unsigned integer. If \texttt{refine} is set to zero, no grid
refinement is performed, and only the adaptive method is applied (see
Section~\ref{sec:method}), bypassing condition~\eqref{eq:refine}. For \texttt{refine}
$\geq 1$, condition~\eqref{eq:refine} is evaluated iteratively up to \texttt{refine}
times. This process progressively generates a more uniformly distributed grid with
increased knot density as \texttt{refine} increases or \texttt{refine\_ns} decreases,
until condition~\eqref{eq:refine} is satisfied or the maximum number of knots, specified
by the input parameter \texttt{max\_nodes}, is reached.

It is worth noting that the \texttt{refine} option does not directly pinpoint regions of
premature convergence, which may result in the addition of unnecessary knots and
increased precision. While increasing $s_\mathrm{conv}$ could also address this issue,
our tests revealed that it often leads to over-sampling and reduced efficiency. By
contrast, the \texttt{refine} option offers a more targeted strategy for improving
accuracy in regions with sharp features, enhancing the overall performance of the
adaptive method.

\section{Testing AutoKnots}
\label{sec:tests}

In this section, we present a series of numerical tests to evaluate the adaptive
methodology described in section~\ref{sec:method}. The tests are divided into two parts.
First, we analyze three distinct functions: a smooth function to demonstrate the basic
application of the adaptive method, a function with sharp features to explore the impact
of the \texttt{refine} option, and a function with varying scales to highlight the
influence of the scale parameter. Second, we assess the method's overall behavior and
stability by applying it to three different sets of functions, each generated randomly
ten million times with variations in specific parameters.

For each test, the approximated functions, $\hat{f}(x)$, are compared to the true
functions, $f(x)$, using a uniformly spaced linear grid with ten thousand (10,000)
nodes. This grid resolution ensures that all relevant properties of the functions are
adequately captured, providing a reliable benchmark for evaluating the performance of
the adaptive method.

The input parameter values remain consistent: the relative tolerance, $\delta$, is set
to $10^{-8}$, and the scale parameter, $\varepsilon$, is set to $0$ unless otherwise
specified. The relative difference between the function, $f(x)$, and the approximated
function, $\hat{f}(x)$, is computed using the following expression, which incorporates
the scale parameter, $\varepsilon$ when present:
\begin{equation} \label{eq:diff_rel_scale}
	\DifRel f(x) = \frac{\DifAbs f(x)}{|f(x)| + \varepsilon},
\end{equation}
where $\DifAbs f(x)$ is defined in eq.~\eqref{eq:abs_error}. The relative difference
between the integral of the function, $I$, and the integral of the approximated
function, $\hat{I}(N)$, across their interval is also estimated:
\begin{equation}\label{eq:diff_integral}
	\DifRel Q = \left| \frac{I - \hat{I}}{I} \right|,
\end{equation}
where $I$ is the exact integral and $\hat{I}$ is the integral of the approximated
function. This parameter is intended to evaluate the overall behavior of the
approximation produced by the adaptive method. It's important to note that this
definition does not incorporate the scale parameter $\varepsilon$ to evaluate its
impact.

The AutoKnots algorithm produces several metrics to assess its performance, each
reflecting a key aspect of the method's accuracy and efficiency. These metrics are:
\begin{itemize}
	\item \textbf{Number of Knots}: Represents the total number of knots in the final
	      approximation, which indicates the computational complexity of the solution.
	\item \textbf{Fails}: The percentage of nodes in the linear grid that
	      exceed the specified relative tolerance. This metric quantifies the extent to which
	      the approximation deviates from the desired accuracy at specific points.
	\item \textbf{Normalized Maximum Relative Difference, $\DifRel f(x)\SubMax /
			      \delta$}: The maximum relative difference between the approximated and true
	      functions across the grid, normalized by the relative tolerance $\delta$. A value
	      larger than one indicates that the method failed to achieve the desired accuracy in
	      some regions, whereas a value significantly smaller than one suggests overestimation
	      of the number of knots.
	\item \textbf{Maximum Relative Difference in Integral, $\DifRel Q$}: Evaluates the
	      accuracy of the integral approximation. This metric is crucial for assessing the
	      overall performance, as the method is not solely an interpolator but also a
	      quadrature estimator.
\end{itemize}
These metrics collectively provide a comprehensive picture of the method's performance,
balancing accuracy and efficiency. By analyzing these outputs, users can identify
whether the algorithm has failed, overestimated, or achieved an optimal knot
distribution.

\subsection{The Adaptive Method Applied to a Smooth Function}
\label{sec:ln}

\begin{figure}
	\centering\includegraphics[width=\columnwidth]{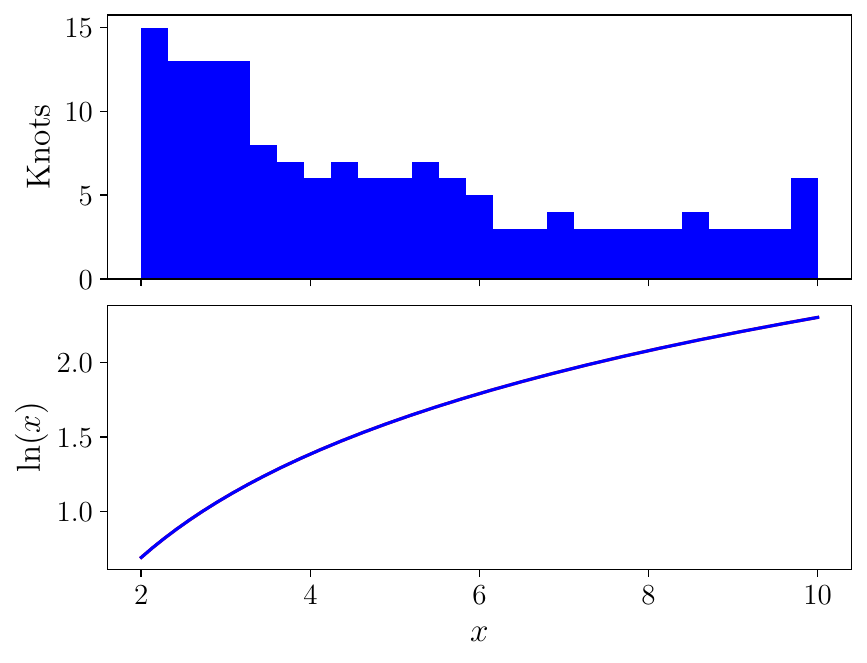}
	\caption{The lower plot depicts the function $f(x) = \ln (x)$, while the upper plot
		illustrates the knot distribution generated by the AutoKnots algorithm.}
	\label{fig:ln}
\end{figure}

The first function analyzed is the natural logarithm, $f(x) = \ln(x)$, evaluated over
the interval $2 \leq x \leq 10$. The adaptive method generates a distribution of knots
that is denser in regions of higher curvature and sparser in regions of lower curvature.
As $x$ increases, the curvature diminishes, and the knot density decreases accordingly,
as illustrated in figure~\ref{fig:ln}.

Interestingly, slight peaks in knot density are observed at the boundaries of the
interval. This behavior is attributed to the reduced precision of cubic spline
interpolation near boundary knots, a known limitation associated with additional
conditions at the edges \citep{SUN2023115039, BEHFOROOZ20068}. Despite this, the
approximated function successfully maintains the desired tolerance throughout the entire
interval. The maximum relative difference, $\DifRel \ln(x)\SubMax / \delta = 0.66$, is
well within the specified tolerance. Additionally, the function meets all precision
requirements, as detailed in the second row of table~\ref{tb:single}.

For comparison, using eq.~\eqref{eq:error_estimate_knots} with an absolute difference
matching the requirement of condition~\eqref{eq:cond1}, $\DifAbs \ln(x) / \delta
	\approx 0.7$ for $x=2$, results in approximately 230 evenly distributed knots. The value at $x =
	2$ was chosen because the fourth derivative of $f(x) = \ln(x)$ is maximized at this
point, making it the most stringent region for the error condition. In contrast, the
adaptive method achieves the same precision with significantly fewer knots.

\begin{table}
	\centering
	\begin{tabular}{ccccccc}
		\hline
		$f(x)$       & Knots & Fails  & $\Delta^{\mathrm{r}}_{\mathrm{max}} / \delta$ & $\Delta^{\mathrm{a}}_{\mathrm{max}} / \delta$ & $\Delta^{\mathrm{r}} Q / \delta$ \\ \hline
		$\ln (x)$    & 153   & 0.00\% & 0.66                                          & 0.53                                          & 0.013                            \\ \hline
		\multirow{2}{*}
		{$I_{d}(x)$} & 927   & 1.24\% & 5.90                                          & 0.71                                          & 0.038                            \\
		             & 945   & 0.00\% & 0.43                                          & 0.06                                          & 0.002                            \\ \hline
		\multirow{4}{*}
		{$N(x)$}     & 2249  & 0.00\% & 0.17                                          & 0.08                                          & 0.002                            \\
		             & 457   & 0.42\% & 0.12                                          & 1.27                                          & 0.023                            \\
		             & 313   & 0.00\% & 0.13                                          & 1.33                                          & 0.120                            \\
		             & 169   & 0.00\% & 0.68                                          & 68.1                                          & 11.10                            \\
		\hline
	\end{tabular}
	\caption{ Numerical results for the AutoKnots method applied to the three test
		functions: $\ln(x)$ (section~\ref{sec:ln}), $I_{d}(x)$
		(section~\ref{sec:test_refine}), and $N(x)$ (section~\ref{sec:test_scale}). The second
		column reports the total number of knots generated. The third column indicates
		the percentage of grid nodes that fail to meet the convergence criteria. The
		fourth and fifth columns present the maximum relative and absolute differences,
		respectively, across all grid nodes. The sixth column shows the integral
		relative difference, as defined in eq.~\eqref{eq:diff_integral}. For $I_{d}(x)$,
		the first row corresponds to results obtained using the adaptive method alone,
		while the second row includes results with \texttt{refine} set to 1 and
		\texttt{refine\_ns} set to 2.5. For $N(x)$, results are shown for the adaptive
		method with scale parameters $\varepsilon = 0$, 1, 10, and 100, listed
		sequentially from top to bottom.}
	\label{tb:single}
\end{table}

\begin{figure}
	\centering\includegraphics[width=\columnwidth]{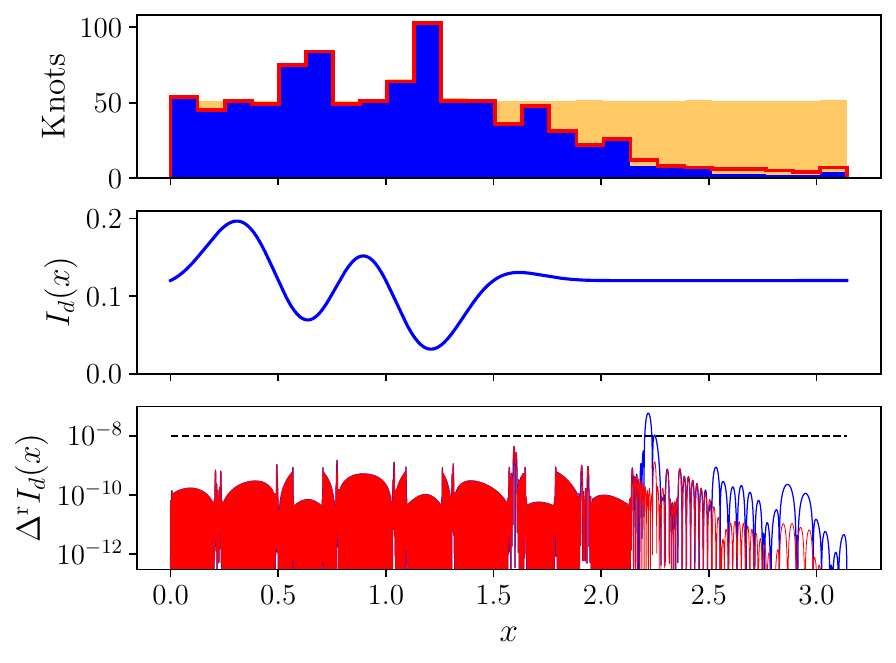}
	\caption{The function $I_{d}(x)$, defined in eq.~\eqref{eq:Idais2018}, is depicted
		in the middle plot. The bottom plot shows the relative difference, $\DifRel
			I_{d}(x)$, at each grid node, with the black dashed line representing the
		convergence threshold. The top plot illustrates the knot distribution: results
		from the adaptive method are represented by the blue line and blue-filled
		histogram, those with \texttt{refine} = 1 and \texttt{refine\_ns} = 2.5 are
		depicted by the red line and red step histogram, and results with
		\texttt{refine} approaching infinity and \texttt{refine\_ns} = 1 are shown by
		the orange-filled histogram.}
	\label{fig:idais}
\end{figure}

\subsection{Testing the \texttt{refine} option}
\label{sec:test_refine}

To evaluate the effectiveness of the \texttt{refine} option, we analyze a function with
contrasting behaviors: rapid oscillations in one region and a nearly constant plateau in
another. This provides a challenging test case for adaptive methods, particularly in
identifying and refining problematic regions.

The function under consideration is given by
\begin{equation} \label{eq:Idais2018}
	I_{d}(x) = 0.12 + 0.25 \exp \left[ -4 \left( x - \pi/4 \right)^2 \right]
	\cos \left( 2x \right) \sin \left( 2 \pi x \right),
\end{equation}
and is evaluated in the interval $0 \leq x \leq \pi$. This function, previously used by
\cite{idais2019optimal} to test a Multi-Objective-Genetic Algorithm (MOGA) for knot
placement in cubic spline interpolation, combines high-frequency oscillations near the
start of the domain with a smooth plateau in the latter half, as shown in the middle
plot of figure~\ref{fig:idais}.

Figure~\ref{fig:idais} illustrates results obtained with three configurations of the
input parameters:
\begin{enumerate}
	\item Baseline Adaptive Method: The adaptive method alone is applied,
	      represented by the blue line/filled histogram. This results in a higher density of
	      knots in the oscillatory region and fewer in the plateau.
	\item Limited Refinement: The AutoKnots algorithm is enhanced with the
	      \texttt{refine} option set to 1 and \texttt{refine\_ns} set to 2.5, shown as a red
	      line/step histogram. Values of \texttt{refine\_ns} up to 5 yield the same outcome,
	      while higher values effectively disable the refinement process.
	\item Aggressive Refinement: The \texttt{refine} parameter is set to a very high
	      value, with \texttt{refine\_ns} set to 1, depicted by the orange-filled histogram.
	      When pushed to the extreme, the refinement process minimizes the variance of the
	      interval widths, effectively making all intervals as small as the smallest one
	      before the refinement starts.
\end{enumerate}
These configurations demonstrate how the \texttt{refine} option influences knot
distribution, improving accuracy in complicated regions while balancing efficiency.

The relative difference, $\DifRel I_{d}(x)$, is shown in the bottom plot of
figure~\ref{fig:idais}, while numerical results for the first two evaluations are
presented in the third and fourth rows of table~\ref{tb:single}, respectively. The
adaptive method fails to meet the specified tolerance within a narrow segment near the
transition to the plateau, with 1.24\% of the grid nodes exceeding the desired
tolerance. The fourth and sixth columns of table~\ref{tb:single} show the maximum
relative difference on the linear grid, $\DifRel I_{d}(x)\SubMax / \delta = 5.9$, which
is within an order of magnitude of the convergence criterion, and the integral relative
difference, $\DifRel Q/\delta = 0.0378$, which falls below the convergence criterion.
These results demonstrate that the adaptive method provides a precise approximation of
the true function.

To achieve not only accuracy but also compliance with the convergence criteria across
the entire interval, the \texttt{refine} option becomes useful. This option is
specifically designed to address situations where small deviations from the required
tolerance must be minimized. By probing intervals with widths larger than the average,
the \texttt{refine} option can identify and correct these regions. The top plot of
figure~\ref{fig:idais} shows the distribution of knots, illustrating that setting
\texttt{refine} to 1 with \texttt{refine\_ns} set to 2.5 increases the knot density
within the plateau region and extends the refinement to all nodes that fail to meet the
tolerance. However, this also results in the introduction of unnecessary knots for $x >
	2.5$. As shown in the first column of table~\ref{tb:single}, this configuration leads to
a 2\% increase in the total number of knots, which in turn reduces the maximum relative
difference to $4.3 \times 10^{-9}$, well within the desired tolerance of $10^{-8}$.

When the number of refinements \texttt{refine} is large and \texttt{refine\_ns} is set
to 1, the knot distribution becomes nearly uniform across the second half of the
function, targeting all intervals that meet condition~\eqref{eq:refine}. This analysis
highlights the effect of excessive refinement values on knot distribution. However, such
an approach is not recommended, as it results in oversampling, making all intervals as
small as the smallest interval before refinement starts. Consequently, numerical results
for this configuration are omitted from table~\ref{tb:single}, and its relative
difference is not included in figure~\ref{fig:idais}.

\begin{figure}
	\centering\includegraphics[width=\columnwidth]{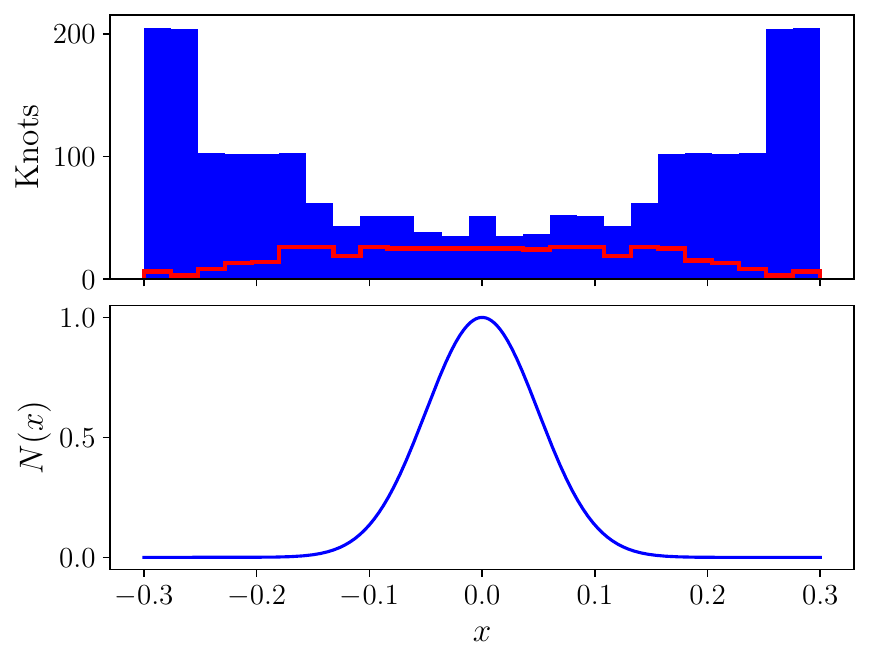}
	\caption{The function $N(x)$ given by eq.~\eqref{eq:gaussian}
		is depicted in the bottom plot. The top plot displays the distribution of knots,
		where the blue-filled histogram represents the adaptive method with $\varepsilon
			= 0$, and the red step histogram represents $\varepsilon = 1$.}
	\label{fig:gaussian}
\end{figure}

\subsection{Scale Parameter}
\label{sec:test_scale}

The third test case illustrates the role of the scale parameter, $\varepsilon$, in
controlling interpolation accuracy. The scale parameter serves two primary purposes: (i)
it defines an absolute tolerance for functions passing through zero, ensuring the
adaptive process terminates even when $f(x_0) = 0$ for $x_0 \in [a,b]$, and (ii) it
allows users to deprioritize regions where the function magnitude is negligible when
compared to the scale parameter. This is particularly useful for functions with
significant variations in magnitude, as it reduces the number of knots in regions where
the function is small, focusing computational resources on areas with more significant
values.

To demonstrate these effects, we employ the Gaussian function:
\begin{equation}\label{eq:gaussian}
	N(x) = \exp{\left[-\frac{1}{2}\left(\frac{x}{0.05}\right)^2\right]},
\end{equation}
evaluated over the interval $-0.3 \leq x \leq 0.3$.

The bottom plot in figure~\ref{fig:gaussian} shows $N(x)$, while the top plot displays the
knot distributions for $\varepsilon=0$ (blue-filled histogram) and $\varepsilon=1$ (red
step histogram). When $\varepsilon=0$, the number of knots increases rapidly as $|N(x)|$
approaches zero, consistent with the behavior described in section~\ref{sec:method}.
Numerical results for $\varepsilon=0$, 1, 10, and 100 are summarized in rows five
through eight of table~\ref{tb:single}. As $\varepsilon$ increases, the number of knots
decreases significantly, reflecting reduced sensitivity to regions where $|N(x)| \ll
	\varepsilon$.

For the scale parameter $\varepsilon=0$, the maximum relative and absolute differences
are $\DifRel N(x)\SubMax/\delta = 0.17$ and $\DifAbs N(x)\SubMax/\delta = 0.08$,
respectively, indicating that the required accuracy is achieved. In this case, the tails
of the function are accurately approximated, as the scale parameter is not introduced.
The integral relative difference is $\DifRel Q/\delta = 0.002$, indicating an
over-sampled approximation. This is expected, as the method is designed to ensure
accuracy in the function tail, even when $N(x)$ is negligible.

For $\varepsilon=1$, the maximum relative difference remains consistent with the
absolute tolerance, $\DifRel N(x)\SubMax/\delta = 0.12$. In regions where $N(x) \ll 1$,
particularly near the tails of the function, the error threshold effectively becomes
$\DifAbs N(x)\SubMax < \delta(\varepsilon + |N(x)|) \approx 10^{-8}$. For instance, if
$N(x_0) = 10^{-9}$ and $\hat{N}(x_0) = 2 \times 10^{-9}$, the relative error with
$\varepsilon=0$ would be $\DifRel N(x_0) = 1$, prompting the AutoKnots method to add more
knots to meet the convergence criteria. In contrast, with $\varepsilon=1$, the same
error would result in $\DifRel N(x_0) \approx 10^{-9}$, which falls well within the
tolerance, and the adaptive method would not increase the number of knots. This
demonstrates how $\varepsilon$ prevents over-sampling in regions where $N(x)$ is
negligible when compared to the scale parameter. Naturally, the absolute
difference is not necessarily smaller than one, in this case, $\DifAbs N(x)\SubMax/\delta
	= 1.27$. This is consistent with the absolute tolerance threshold.

To illustrate the effects of choosing $\varepsilon$ larger than the maximum value of
$N(x)$, we analyze cases where $\varepsilon = 10$ and $\varepsilon = 100$. When
$\varepsilon > \max N(x)$, the relative tolerance $\delta$ is effectively reduced,
leading to a global decrease in interpolation precision. For $\varepsilon = 10$, this
reduction is modest, as the value is only one order of magnitude larger than $\max
	N(x)$, resulting in a reasonably accurate approximation. However, for $\varepsilon =
	100$, the effects are more pronounced, with $\DifAbs N(x)\SubMax/\delta = 68.1$ and
$\DifRel Q/\delta = 11.10$, indicating an approximation that fails to meet the desired
tolerance anywhere within the domain, see the last row of table~\ref{tb:single}. This
highlights the importance of carefully selecting $\varepsilon$ to balance precision and
efficiency.

The choice of $\varepsilon$ should align with the scale of the smallest function values
of interest to achieve the best results. For $N(x)$, while $|N(x)|\SubMax = 1$, the
optimal outcome was obtained with $\varepsilon \approx 1$, highlighting that both the
function's maximum value and the desired balance between accuracy and efficiency play a
role in determining the appropriate scale parameter.

\begin{figure}
	\centering\includegraphics[width=\columnwidth]{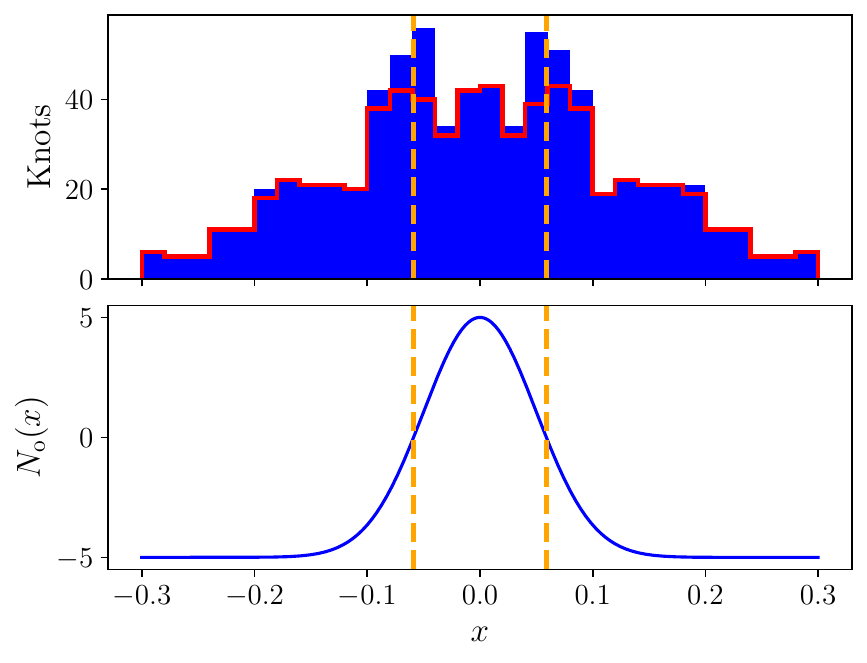} \caption{The
		function $N_{\mathrm{o}}(x) = 10 N(x) - 5$ is depicted in the bottom plot. The top
		plot displays the distribution of knots: the blue-filled histogram represents the
		adaptive method with $\varepsilon = 0$, and the red step histogram represents
		$\varepsilon = 1$. The vertical orange dashed lines in both plots indicate where the
		function crosses the abscissa.}
	\label{fig:gaussian_offset}
\end{figure}

To assess the impact of a function crossing the abscissa, we studied the function
$N_{\mathrm{o}}(x) = 10 N(x) - 5$, evaluated within the same interval. The amplitude and
offset values were chosen to minimize the impact of the scale parameter beyond the
abscissa crossing. Figure~\ref{fig:gaussian_offset} illustrates the function and the
distribution of knots for $\varepsilon = 0$ and $\varepsilon = 1$. The impact of the
scale parameter is noticeable only where the function crosses the abscissa. The number
of knots is similar, decreasing by 9\% from 733 to 669, and both approximated functions
share the same maximum absolute difference. This demonstrates that the AutoKnots algorithm
is not highly sensitive to this type of behavior.

For comparison, we also analyzed the $\ln(x)$ and $I_{d}(x)$ functions (see
section~\ref{sec:ln} and section~\ref{sec:test_refine}) using $\varepsilon = 1$. For $\ln(x)$,
the number of knots decreased slightly, from 153 to 135, with negligible changes in all
other parameters. The function $I_{d}(x)$ exhibited a more pronounced effect, with the
number of knots reduced significantly from 927 to 539, while the maximum absolute
difference remained nearly unchanged. This variation arises because $I_{d}(x)\SubMax
	\approx 0.2$, making the absolute tolerance dominant in the optimization process,
whereas $\ln(x=2) \approx 0.7$, causing $\varepsilon = 1$ to affect only the beginning
of its interval. In practice, determining the optimal $\varepsilon$ value is often
challenging, as it depends on the function's behavior and may involve parameters unknown
\textit{a priori}. Hence, it is generally advisable to set $\varepsilon = 0$ or choose
the smallest value that still captures the function's features accurately.

These results underscore the difficulty of selecting an optimal scale parameter for a
given function. There is no universal method to determine the ideal $\varepsilon$ value,
and its choice typically requires trial and error, as illustrated with the Gaussian
function. Choosing $\varepsilon = 0$ is a conservative approach that guarantees a fit
within the desired tolerance but can result in over-sampling under certain conditions.
It is worth noting that $\varepsilon = 0$ corresponds to setting the absolute tolerance
to zero in other numerical techniques, such as quadrature methods, thereby demanding a
similar balance between precision and efficiency.

\begin{figure}
	\centering\includegraphics[width=\columnwidth]{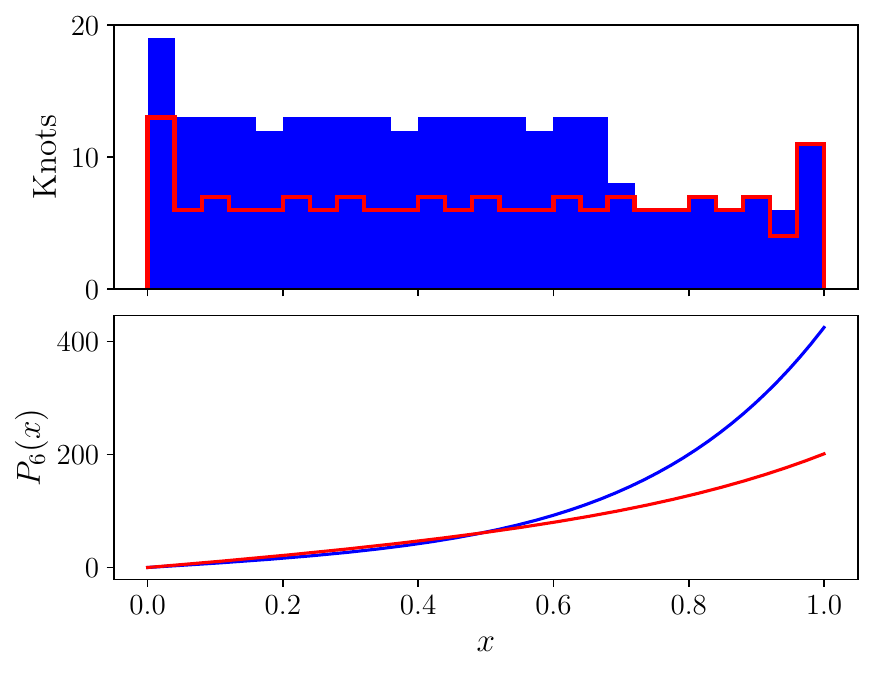}
	\caption{The lower plot illustrates two realizations of the function $P_{6}(x)$,
		defined by eq.~\eqref{eq:P6}, while the upper plot displays their respective knot
		distributions generated by the adaptive method.}
	\label{fig:P6_example}
\end{figure}

\begin{figure}
	\centering\includegraphics[width=\columnwidth]{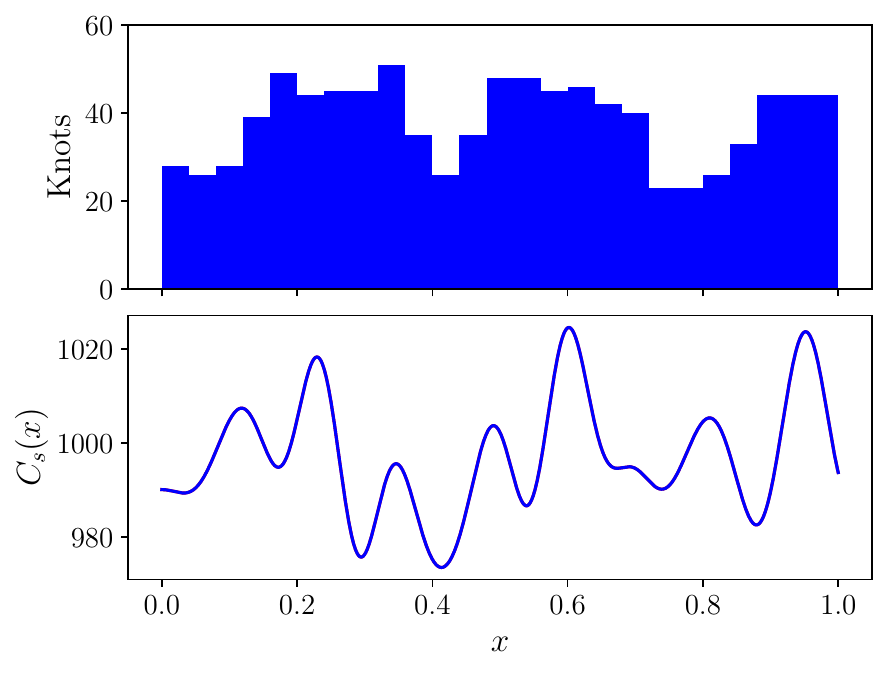}
	\caption{The lower plot illustrates a single realization of the function $C_{s}(x)$
		defined by eq.~\eqref{eq:cos_sum}, while the upper plot shows the knot distribution
		generated by the adaptive method.}
	\label{fig:Cs_example}
\end{figure}

\begin{figure}
	\centering\includegraphics[width=\columnwidth]{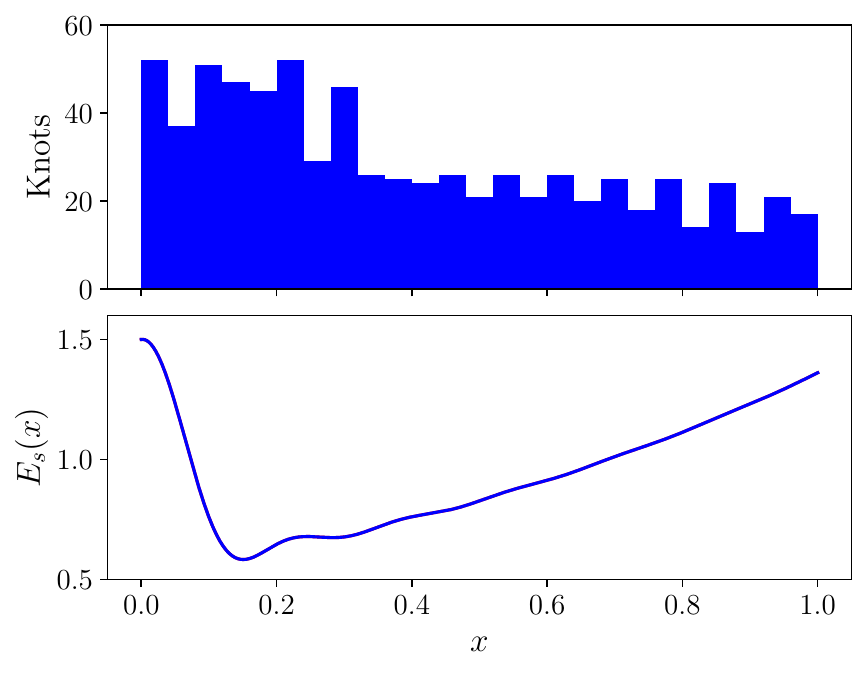} \caption{The lower plot
		illustrates a single realization of the function $E_{s}(x)$ defined by
		eq.~\eqref{eq:exp_sin2_sim}, while the upper plot shows the knot distribution
		generated by the adaptive method.}
	\label{fig:E_s}
\end{figure}

\begin{table}
	\centering
	\begin{tabular}{cccccc}
		\hline
		$f(x)$     & Refine & Knots          & Fails   & $\frac{\DifRel f(x)_{\mathrm{max}}}{\delta}$ & MAX             \\ \hline
		$P_{6}(x)$ & (0, 0) & 260 $\pm$ 51.9 & 0.3\%   & 0.7 $\pm$ 9                                  & 500             \\
		$P_{6}(x)$ & (1, 5) & 260 $\pm$ 52.2 & 0\%     & 0.2 $\pm$ 0.1                                & 0.9             \\
		$P_{6}(x)$ & (1, 1) & 291 $\pm$ 64.4 & 0\%     & 0.2 $\pm$ 0.08                               & 0.4             \\
		\hline
		$C_{s}(x)$ & (0, 0) & 1070 $\pm$ 218 & 40\%    & 9 $\pm$ $4\times10^3$                        & $7 \times 10^6$ \\
		$C_{s}(x)$ & (1, 5) & 1080 $\pm$ 220 & 4\%     & 0.4 $\pm$ 0.4                                & 50              \\
		$C_{s}(x)$ & (1, 1) & 1180 $\pm$ 251 & 0.002\% & 0.07 $\pm$ 0.04                              & 50              \\
		\hline
		$E_{s}(x)$ & (0, 0) & 625 $\pm$ 186  & 20\%    & 3 $\pm$ 300                                  & $3\times 10^3$  \\
		$E_{s}(x)$ & (1, 5) & 627 $\pm$ 186  & 4\%     & 0.4 $\pm$ 1                                  & $30$            \\
		$E_{s}(x)$ & (1, 1) & 689 $\pm$ 188  & 0.002\% & 0.1 $\pm$ 0.09                               & $20$            \\
		\hline
	\end{tabular}
	\caption{Numerical results for the adaptive method applied to the functions $P_6(x)$
		(eq.~\eqref{eq:P6}), $C_s(x)$ (eq.~\eqref{eq:cos_sum}), and $E_s(x)$
		(eq.~\eqref{eq:exp_sin2_sim}). The second column specifies the refinement
		applied. The third column presents the mean and standard deviation of the number
		of knots generated across all realizations. The fourth column reports the
		percentage of realizations that failed to meet the convergence criteria (Fails).
		The fifth column provides the mean and standard deviation of the maximum
		relative error for all realizations. Finally, the last column shows the absolute
		maximum relative error across all realizations. }
	\label{tb:func_sim_method}
\end{table}

\subsection{Statistical Performance Evaluation}
\label{sec:test_additional}

In this section, we evaluate the statistical performance of the AutoKnots method by
applying it to a broad set of parametrized functions with randomly sampled parameters.
The goal is to test the algorithm's robustness and versatility across a diverse range of
functional forms, simulating scenarios where function parameters are inferred from data,
such as in optimization techniques like maximum likelihood estimation or Markov Chain
Monte Carlo. To this end, we generate ten million ($10^7$) random samples from three
distinct sets of functions and compare the fitted results to the true functions using a
uniform linear grid of ten thousand nodes for each realization. All analyses are
performed over the interval $0 \leq x \leq 1$.

The evaluation is purely numerical, relying solely on the data obtained from the
realizations without invoking any theoretical probability distribution for the
parameters. It is assumed that the dataset of ten million realizations is sufficiently
large to approximate the true distribution, enabling a reliable assessment of the
method's performance.

To evaluate the method's performance across a wide range of functions, we focus on the
following metrics:
\begin{itemize}
	\item \textbf{Knots}: For each realization, we apply AutoKnots to the function and
	      generate a distribution of knots. We then compute the mean and standard
	      deviation of the number of knots across all $10^7$ realizations for each
	      function.
	\item \textbf{Fails}: For each realization, we compute the approximation and
	      evaluate it on the grid, comparing the approximation to the original function.
	      If the approximation at any grid point fails to recover the original function
	      value within the desired tolerance, that realization is marked as failed. The
	      reported percentage of failures (Fails) is the total number of failed
	      realizations divided by $10^7$.
	\item \textbf{Maximum Relative Difference}: For each grid point in a realization, we
	      compute the relative difference as ${\DifRel f(x)_{\mathrm{max}}}/{\delta}$.
	      For each realization, we select the maximum relative difference across all
	      grid points. Using these $10^7$ maximum values, we compute the mean and
	      standard deviation, as well as the maximum value across all realizations
	      (MAX).
\end{itemize}
The adaptive method is tested on three sets of functions, each with distinct
characteristics. For each set, we evaluate the method under three configurations: no
refinement $(0,0)$, one refinement with $\texttt{refine\_ns} = 5$ $(1,5)$, and one
refinement with $\texttt{refine\_ns} = 1$ $(1,1)$.

The initial function is a sixth-order polynomial. This function set was specifically
designed to assess the performance of the adaptive method on smooth functions. Its
expression is as follows:
\begin{equation}\label{eq:P6}
	P_{6}(x) = \sum_{i=1}^{6} A_i \, x^i \, .
\end{equation}
In each realization, the parameters $A_i$ are randomly drawn from a uniform distribution
within $1 \leq A_i \leq 100$. It's noteworthy that all instances of $P_{6}(x)$ have a
value of zero at $x = 0$ and are increasing positive functions.
Figure~\ref{fig:P6_example} illustrates two instances of $P_{6}(x)$, showcased in the
lower plot, along with the corresponding distributions of knots depicted in the upper
plot.

The second set of functions aims to assess the performance of the AutoKnots algorithm under
conditions of high variability, defined as:
\begin{equation}\label{eq:cos_sum}
	C_{s}(x) = \sum_{i=1}^{29} B_{i} \cos \left( 2 \pi \nu_i  x \right) + 10^{3} \, .
\end{equation}
The amplitudes $B_i$ and frequencies $\nu_i$ are randomly sampled from a normal
distribution with a mean of zero and a standard deviation of 5. The constant factor has
an equivalent effect to adding the scale parameter with the same value, $\varepsilon =
	1000$, influencing the absolute difference threshold as $\DifAbs C_{s}(x)\SubMax \approx
	1000 \, \delta = 10^{-5}$. Figure~\ref{fig:Cs_example} illustrates an instance of such
realization, showcasing $C_{s}(x)$ in the lower plot and presenting the corresponding
distribution of knots in the upper plot.

The third and final set of functions is given by:
\begin{equation}\label{eq:exp_sin2_sim}
	E_{s}(x) = 0.5 \, \mathrm{e}^{\alpha x} + \left[\frac{\sin(\beta x)}{\beta x} \right]^{2} \, ,
\end{equation}
The parameters $\alpha$ and $\beta$ are randomly sampled from uniform distributions
within the intervals $0.1 \leq \alpha \leq 2$ and $5 \leq \beta \leq 30$.
Figure~\ref{fig:E_s} displays an example of a realization with $\alpha = 1$ and $\beta =
	20$. This function serves as an intermediary between the smoothness of $P_{6}(x)$ and
the high variability of $C_{s}(x)$.

The numerical results of this analysis, summarized in table~\ref{tb:func_sim_method},
highlight the relationship between function characteristics and the number of knots
generated. As expected, the smooth function $P_{6}(x)$ requires fewer knots compared to
the highly variable $C_{s}(x)$. The function $E_{s}(x)$, with intermediate smoothness
and variability, falls between these two extremes.

\begin{figure}
	\centering\includegraphics[width=\columnwidth]{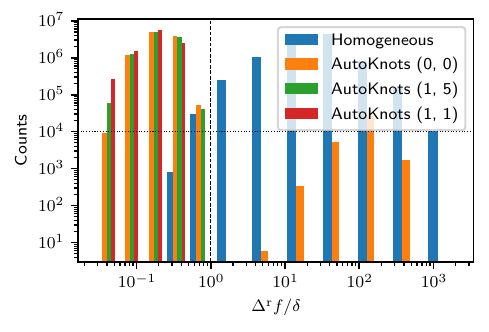}
	\caption{Histograms showing the distribution of the maximum relative error for
		$P_{6}(x)$. Bins to the right of 1 (vertical dashed line) correspond to failed
		realizations, while bins below the dotted line represent less than 0.1\% of the
		total realizations. AutoKnots without refinement exhibits a second mode to the right
		of 1, indicating feature loss in some realizations. The refinement process shifts
		more realizations to the left, improving accuracy. A homogeneous knot distribution
		(with the same number of knots) performs significantly worse, with most bins to the
		right of 1. See figure~\ref{fig:cosine_hist} and figure~\ref{fig:exp_cos_hist} for
		similar analyses of $C_{s}(x)$ and $E_{s}(x)$.}
	\label{fig:P6_hist}
\end{figure}

\subsubsection{Smooth Function $P_{6}(x)$}

For $P_{6}(x)$, just 0.3\% of the realizations fail to meet the tolerance when no refinement
is applied. However, within these few cases, the maximum relative
difference reaches 500, or about three orders of magnitude above the tolerance ($500
	\times 10^{-8} = 5 \times 10^{-6}$). When one refinement with $\texttt{refine\_ns} = 5$
is applied, the failure rate drops to zero, and the maximum relative difference reduces
to $0.2 \pm 0.1$, with an absolute maximum of 0.9. This demonstrates excellent
performance, achieving results very close to the desired tolerance without generating
unnecessary knots. Increasing the refinement to $\texttt{refine\_ns} = 1$ results in a
12\% increase in the number of knots. While the failure rate remains at zero, the
maximum relative difference improves slightly to $0.2 \pm 0.08$, with a reduced absolute
maximum of 0.4. This configuration provides the best results, with nearly all
realizations meeting the desired tolerance.

In figure~\ref{fig:P6_hist}, the plot illustrates the distribution of the maximum relative
error for $P_{6}(x)$, highlighting a minor mode near $\DifRel P_{6}(x) \approx 10^{-6}$,
which corresponds to realizations that fail to meet the required tolerance. These
realizations are better approximated when refinement is applied, demonstrating how the
refinement process effectively identifies and corrects problematic regions. One last
interesting aspect is that, for all applications of the algorithm, the approximations do
not exceed the required tolerance by much more than one order of magnitude. This shows
that the method does not use excessive knots, achieving efficient approximation without
unnecessary complexity.

\begin{figure}
	\centering\includegraphics[width=\columnwidth]{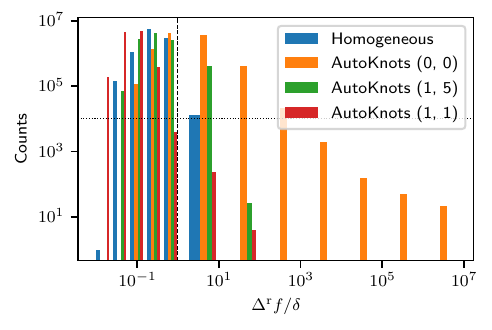}
	\caption{Histograms for the maximum relative error of $C_{s}(x)$, with the same
		conventions as figure~\ref{fig:P6_hist}. Without refinement, AutoKnots fail more
		often, showing a second mode to the right of 1. Refinement significantly improves
		performance, shifting realizations left. The homogeneous knot distribution performs
		better than AutoKnots without refinement but worse than AutoKnots with refinement,
		as seen by its higher failure rate and larger first bin to the right of 1.}
	\label{fig:cosine_hist}
\end{figure}

\subsubsection{Highly Variable Function $C_{s}(x)$}

Now, let us discuss the results for $C_{s}(x)$. When no refinement is applied, the
adaptive method generates several knots, with a mean of 1070 and a standard
deviation of 218. The failure rate is 40\%, with a maximum relative difference of $9 \pm
	4 \times 10^3$ and an absolute maximum of $7 \times 10^6$. These results indicate that
the AutoKnots method with no refinement struggles to approximate the function accurately, with a
significant proportion of realizations exceeding the desired tolerance.

When one refinement with $\texttt{refine\_ns} = 5$ is applied, the number of knots
remains similar, with a mean of 1080 and a standard deviation of 220. The failure rate
decreases to 4\%, with a maximum relative difference of $0.4 \pm 0.4$ and an absolute
maximum of 50. This configuration demonstrates a substantial improvement, with the
majority of realizations now meeting the desired tolerance.

Increasing the refinement further to $\texttt{refine\_ns} = 1$ results in a 10\% increase
in the number of knots, with a mean of 1180 and a standard deviation of 251. The failure
rate drops dramatically to 0.002\%, with a maximum relative difference of $0.07 \pm 0.04$
and an absolute maximum of 50. This configuration yields the best performance, with
nearly all realizations meeting the desired tolerance.

One notable result for $C_{s}(x)$ is the surprisingly good performance of the
homogeneous knot distribution, as shown in figure~\ref{fig:cosine_hist}. It results in
fewer failed realizations than the adaptive method without refinement and shows a better
maximum relative difference. This is likely due to the periodic nature of the function,
where the homogeneous distribution can capture the oscillations more efficiently than
the adaptive method without refinement. However, since the homogeneous distribution uses
the same number of knots as the adaptive method (derived from AutoKnots), it is not a
fair comparison. In practice, one would need to know the optimal number of knots in
advance.

When compared to AutoKnots with refinement, the homogeneous distribution performs worse.
For example, the first bin to the right of 1 in the homogeneous distribution is
significantly larger than in AutoKnots with $(1, 1)$. This difference is reflected in
the failure rates: 0.002\% for AutoKnots with $(1, 1)$ versus 0.1\% for the homogeneous
distribution.

\begin{figure}
	\centering\includegraphics[width=\columnwidth]{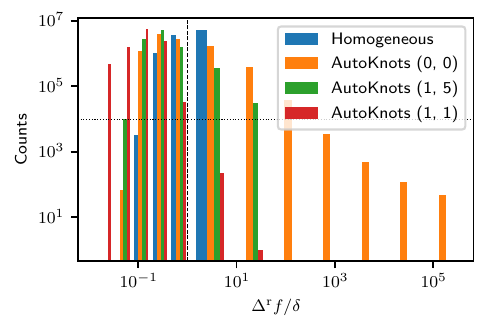}
	\caption{Histograms for the maximum relative error of $E_{s}(x)$, with conventions
		matching figures~\ref{fig:P6_hist} and \ref{fig:cosine_hist}. Without refinement,
		AutoKnots exhibits a significant number of failed realizations and a second mode to
		the right of 1. Refinement reduces the failure rate and improves accuracy. The
		homogeneous knot distribution outperforms AutoKnots without refinement but performs
		worse than AutoKnots with refinement, showing higher failure rates and larger error
		bins.}
	\label{fig:exp_cos_hist}
\end{figure}

\subsubsection{Intermediate Function $E_{s}(x)$}

For $E_{s}(x)$, the adaptive method without refinement uses an average of 625 knots
(standard deviation: 186) and has a failure rate of 20\%. The maximum relative error is
$3 \pm 300$, with an absolute maximum of $3 \times 10^3$. Adding one refinement step
($\texttt{refine\_ns} = 5$) maintains a similar knot count (mean: 627, standard
deviation: 186) but reduces the failure rate to 4\%. The maximum relative error drops
significantly to $0.4 \pm 1$, with an absolute maximum of 30. This shows that refinement
substantially enhances performance, ensuring most realizations meet the tolerance.

Further increasing the refinement ($\texttt{refine\_ns} = 1$) raises the average knot
count by 10\% (mean: 689, standard deviation: 188) and lowers the failure rate to just
0.002\%. The maximum relative error becomes $0.1 \pm 0.09$, with an absolute maximum of
20, yielding the best results with nearly all realizations meeting the tolerance.

Comparing the homogeneous knot distribution with AutoKnots, the homogeneous approach has
a higher failure rate without refinement due to the function's intermediate smoothness
and variability, which the adaptive method captures more effectively. However, the
homogeneous method achieves smaller maximum relative errors in some cases, suggesting
better approximation for specific realizations. Even so, AutoKnots with refinement
outperforms the homogeneous approach across all metrics, as shown in
figure~\ref{fig:exp_cos_hist}.

\subsubsection{Refinement Analysis}

Considering the results for $P_{6}(x)$, $C_{s}(x)$, and $E_{s}(x)$, the refinement
process significantly improves the adaptive method's performance. The number of knots
remains relatively stable, with a slight increase when refinement is applied. For smooth
functions like $P_{6}(x)$, the number of knots is increased by 12\% when refinement is
applied unnecessarily. However, for functions with higher variability, such as
$C_{s}(x)$ and $E_{s}(x)$, the increase is necessary to ensure the desired tolerance.

The refinement process also reduces the failure rate, with the most significant
improvement observed for $C_{s}(x)$, where the failure rate drops from 40\% to 0.002\%
when refinement is applied. The maximum relative difference is also significantly
reduced, with the absolute maximum falling from $7 \times 10^6$ to 50 for $C_{s}(x)$.
This demonstrates the effectiveness of the refinement process in improving the adaptive
method's performance.

Naturally, if the function to be approximated is smooth, the refinement process may not
be necessary, as seen in the case of $P_{6}(x)$. However, for functions with higher
variability, such as $C_{s}(x)$ and $E_{s}(x)$, the refinement process is essential to
ensure the desired tolerance is met. In general, the more conservative approach is to
apply the refinement process, as it ensures the adaptive method performs well across a
wide range of functions while only marginally increasing the number of knots. The
default configuration used in NumCosmo is to apply one refinement with
$\texttt{refine\_ns} = 1$, as this provides the best results across all functions
tested.

\section{Applications in Cosmology}
\label{sec:cosmology}

The  AutoKnots algorithm has been integrated into various
components of NumCosmo to streamline the configuration of cosmological models and
calculations. By automating the choice of knot placement based on convergence criteria
and interpolation error thresholds, AutoKnots reduces the need for manual tuning of
configuration parameters, such as the number of knots. This capability is especially
valuable in complex cosmological codes, where the range of required parameters can be
extensive.

In this section, we discuss the range of applications of AutoKnots within NumCosmo,
highlighting its versatility and effectiveness. We describe specific objects and
functions that use the algorithm and provide detailed results of two applications.
These examples illustrate how AutoKnots simplifies the setup of cosmological analyses
while maintaining high precision and computational efficiency.

\begin{itemize}
	\item \textbf{Map on a spherical shell:} The \texttt{NcmSphereMap}
	      object\footnote{\texttt{NcmSphereMap} is a re-implementation of the
		      \texttt{HEALPix} code\citep{healpix2005}.} provides a set of functions
	      dedicated to pixelating the spherical surface (sky), given a resolution,
	      performing coordinate system transformations, and computing the $a_{lm}$
	      coefficients of a spherical harmonics decomposition. It also calculates the
	      two-point angular functions in real and Fourier spaces: $C(\theta)$ and $C_l$,
	      respectively.\footnote{In particular, the method
		      \texttt{ncm\_sphere\_map\_calc\_Ctheta} is used to compute $C(\theta)$.}
	\item \textbf{Spherical Bessel (SB) function:} NumCosmo provides three methods to
	      compute the SB function: a multi-precision version, an approximation up to the
	      third order of the SB's power series, and one using spline interpolation. The
	      spline is built by calculating the knots from the multi-precision SB
	      function.\footnote{The method for this is \texttt{ncm\_sf\_sbessel\_spline}.}
	\item \textbf{Filtered power spectrum:} This method computes a power spectrum $P(k,
		      z)$ filtered by a window function, where $k$ is the wave-number and $z$ is the
	      redshift. The calculation is optimized by constructing a bi-dimensional spline
	      in $k$ and $z$. The spline in $k$ is derived using the Fast Fourier Transform
	      of $P(k, z)$ with logarithmically spaced points (FFTLog) in the range $(\ln
		      k_i, \ln k_f)$. The method \texttt{ncm\_powspec\_filter\_prepare} is applied
	      to compute the respective spline in $z$ and its integral.
	\item \textbf{Correlation function in 3D:} The \texttt{NcmPowspecCorr3d} object
	      computes the 3D two-point spatial correlation function from the filtered power
	      spectrum. The method used is \texttt{ncm\_powspec\_corr3d\_prepare}.
	\item \textbf{Halo mass function:} The \texttt{NcHaloMassFunction} object provides
	      various functions, such as computing the number density of halos as a function
	      of halo mass $M$ and redshift $z$, i.e., $\dd^2N/\dd z\dd\ln M$, the halo
	      distribution in $z$, $\dd N/\dd z$, and the total number of halos $N$ given
	      mass intervals and $z$. The 2D adaptive approach is used to build the 2D
	      spline in $z$ and $\ln M$ of the halo mass function. \footnote{The method is
		      \texttt{\_nc\_halo\_mass\_function\_generate\_2Dspline\_knots}.}
	\item \textbf{Perturbation:} The \texttt{NcHIPert} object solves the ordinary
	      differential equations (ODE) for perturbations in spatially homogeneous and
	      isotropic cosmologies. A common problem is determining the modes for which the
	      ODEs should be solved to compute the power spectrum. This method offers a
	      natural way to select the best knots for the required precision, a process
	      typically done \textit{a priori} in codes like CAMB. Our approach applies to
	      the WKB approximation. The method used is
	      \texttt{\_nc\_hipert\_wkb\_prepare\_approx}.
	\item \textbf{Recombination:} The \texttt{NcRecomb} object describes the universe's
	      recombination period, specifically the fraction of light elements in various
	      states and the number of free electrons. The adaptive method is used to
	      compute the Seager implementation~\cite{Seager1999}.\footnote{The method is
		      \texttt{\_nc\_recomb\_seager\_prepare}.}
	\item \textbf{Halo density profile:} The \texttt{NcHaloDensityProfile} object
	      implements the matter halo density profile $\rho (r)$, the enclosed mass in a
	      spherical volume given $\rho (r)$, the surface mass density $\Sigma(R)$, and
	      the average surface mass density within a cylinder of radius $R$,
	      $\overline{\Sigma}(< R)$. The last two functions use the adaptive
	      method.\footnote{The methods are:
		      \begin{itemize}
			      \item \texttt{\_nc\_halo\_density\_profile\_prepare\_dl\_2d\_density},
			      \item \texttt{\_nc\_halo\_density\_profile\_prepare\_dl\_cyl\_mass}.
		      \end{itemize}} These methods return the dimensionless surface mass density
	      $\hat{\Sigma}(R)$ and the enclosed mass in an infinite cylinder,
	      $\hat{\overline{\Sigma}}(< R)$, of radius $R$.
\end{itemize}

It is worth noting that the
Firecrown~\footnote{\url{https://github.com/LSSTDESC/firecrown}} and
CLMM~\footnote{\url{https://github.com/LSSTDESC/CLMM}} packages utilize certain NumCosmo
tools that incorporate the AutoKnots algorithm. Firecrown and CLMM are official tools of
the Dark Energy Science Collaboration (DESC) of the LSST. Reference~\cite{CLMM2021}
presents a comparison of the accuracy of various functions related to the matter halo
density profile across CCL, Cluster
Toolkit~\footnote{\url{https://cluster-toolkit.readthedocs.io/en/latest/}}, Colossus,
and NumCosmo.

In the following, we evaluate the performance of the method by computing the
dimensionless quantities $\hat{\Sigma}(R)$ and $\hat{\overline{\Sigma}}(<R)$ for two
different halo density profiles: the Navarro-Frenk-White (NFW) profile~\citep{NFW1996}
and the Hernquist profile~\citep{Hernquist1990}. These profiles were selected because
they both provide analytical expressions for $\hat{\Sigma}(R)$ and
$\hat{\overline{\Sigma}}(R)$, allowing us to compare the results obtained using the
adaptive method with the analytical solutions.

For each value of $R$, we compute the dimensionless surface mass density
$\hat{\Sigma}(R)$ and the average surface mass density within a cylinder of radius $R$,
$\hat{\overline{\Sigma}}(<R)$, for both the NFW and Hernquist profiles. These
calculations involve integrating the matter density profiles along the line of sight for
$\hat{\Sigma}(R)$ and within the cylinder for $\hat{\overline{\Sigma}}(<R)$. Since each
new value of $R$ requires fresh integrations, we utilize the AutoKnots algorithm to
dynamically generate $R$ values and their corresponding interpolated functions. This
approach minimizes the number of knots and consequently the computational cost,
while ensuring the desired precision.

\begin{figure}
	\centering\includegraphics[width=\columnwidth]{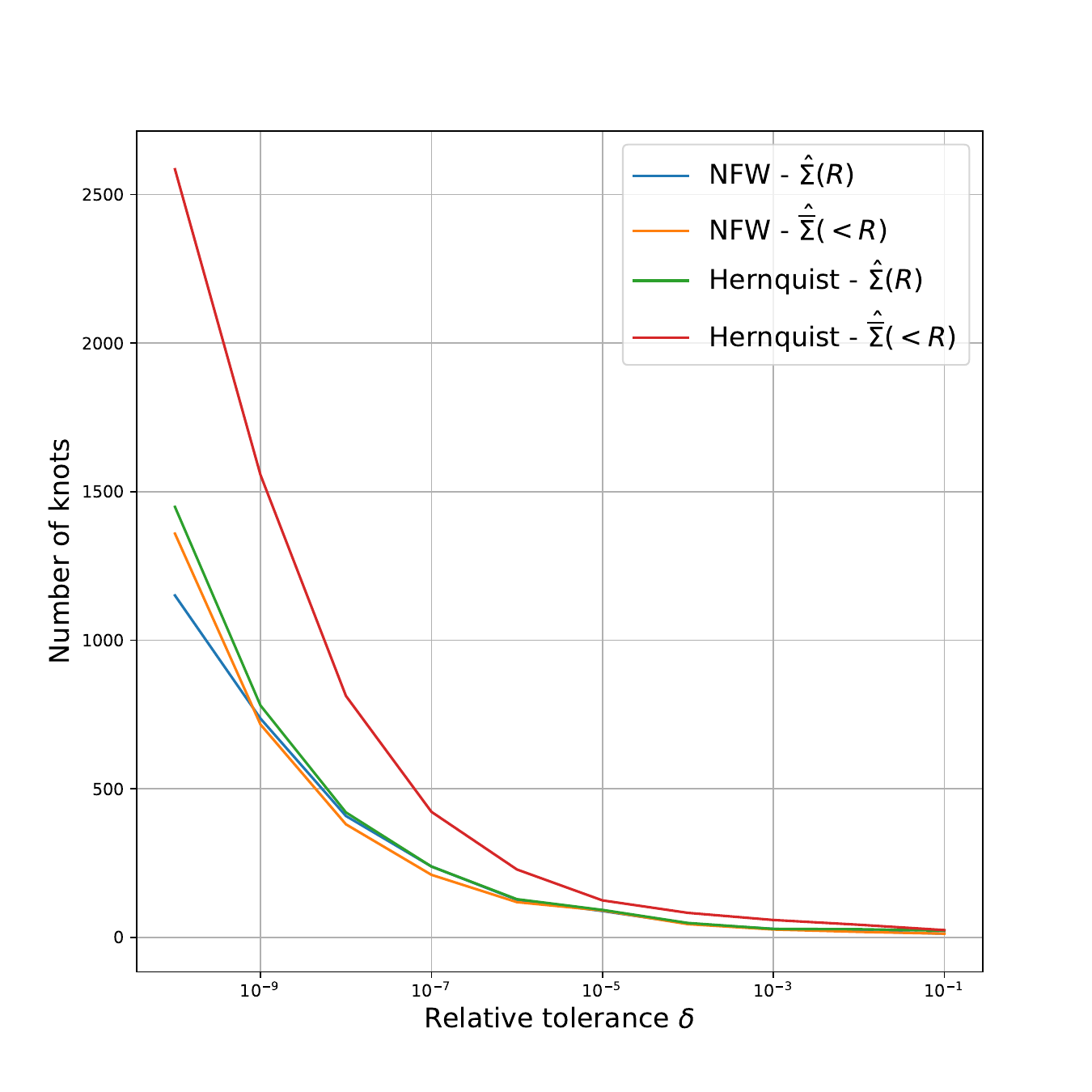} \caption{Number of
		knots as a function of the relative tolerance $\delta$ computed in the interval
		$\delta \in [10^{-10}, 10^{-1}]$ for the four cases: $\hat{\Sigma}(R)$ and
		$\hat{\overline{\Sigma}}(<R)$ using NFW (blue and orange curves) and Hernquist
		(green and red curves) profiles. The plot illustrates how the number of knots
		increases with decreasing tolerance for both profiles.}
	\label{fig:Number_knots_cosmo}
\end{figure}

We calculate $\hat{\Sigma}(R)$ and $\hat{\overline{\Sigma}}(R)$ with the relative
tolerance $\delta$ ranging from $10^{-10}$ to $10^{-1}$. For these functions, we apply
the default refinement option. Additionally, as they are positive-definite, we set the
scale parameter $\epsilon = 0$. In figure~\ref{fig:Number_knots_cosmo}, we present the
number of knots required and how it increases with the relative tolerance for the four
cases considered. It is worth noting that the number of knots increases more rapidly
when $\delta > 10^{-6}$.

\begin{figure}
	\centering\includegraphics[width=\columnwidth]{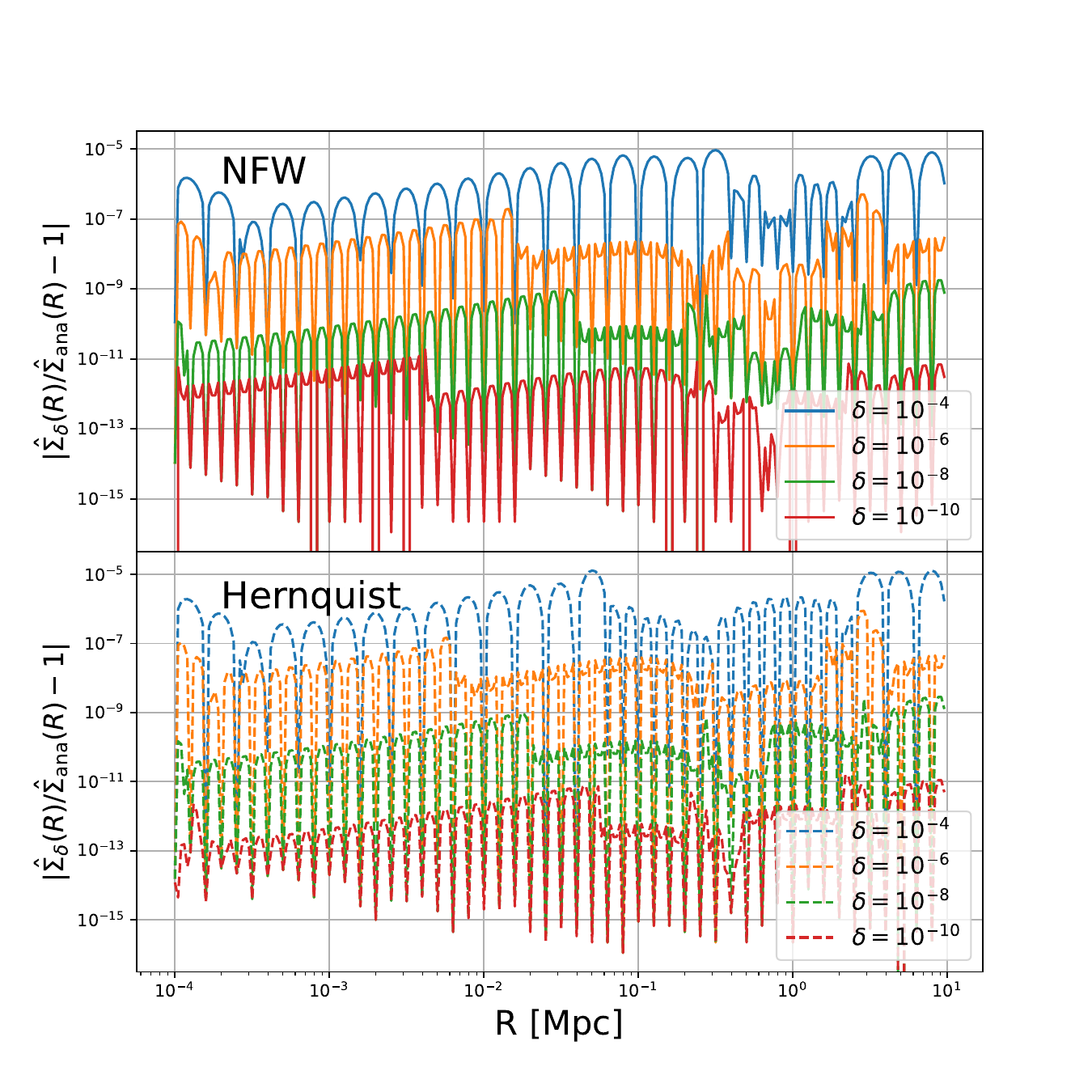}
	\caption{The relative difference between the dimensionless surface mass density
		$\hat{\Sigma}(R)$ computed using spline interpolation and analytically for the NFW
		matter density profile (upper panel) and the Hernquist profile (lower panel). In
		both cases, the splines were built requiring a relative tolerance $\delta$ equal to
		$10^{-4}$, $10^{-6}$, $10^{-8}$ and $10^{-10}$ corresponding to the blue, orange,
		green and red curves, respectively.}
	\label{fig:Sigma}
\end{figure}

\begin{figure}
	\centering\includegraphics[width=\columnwidth]{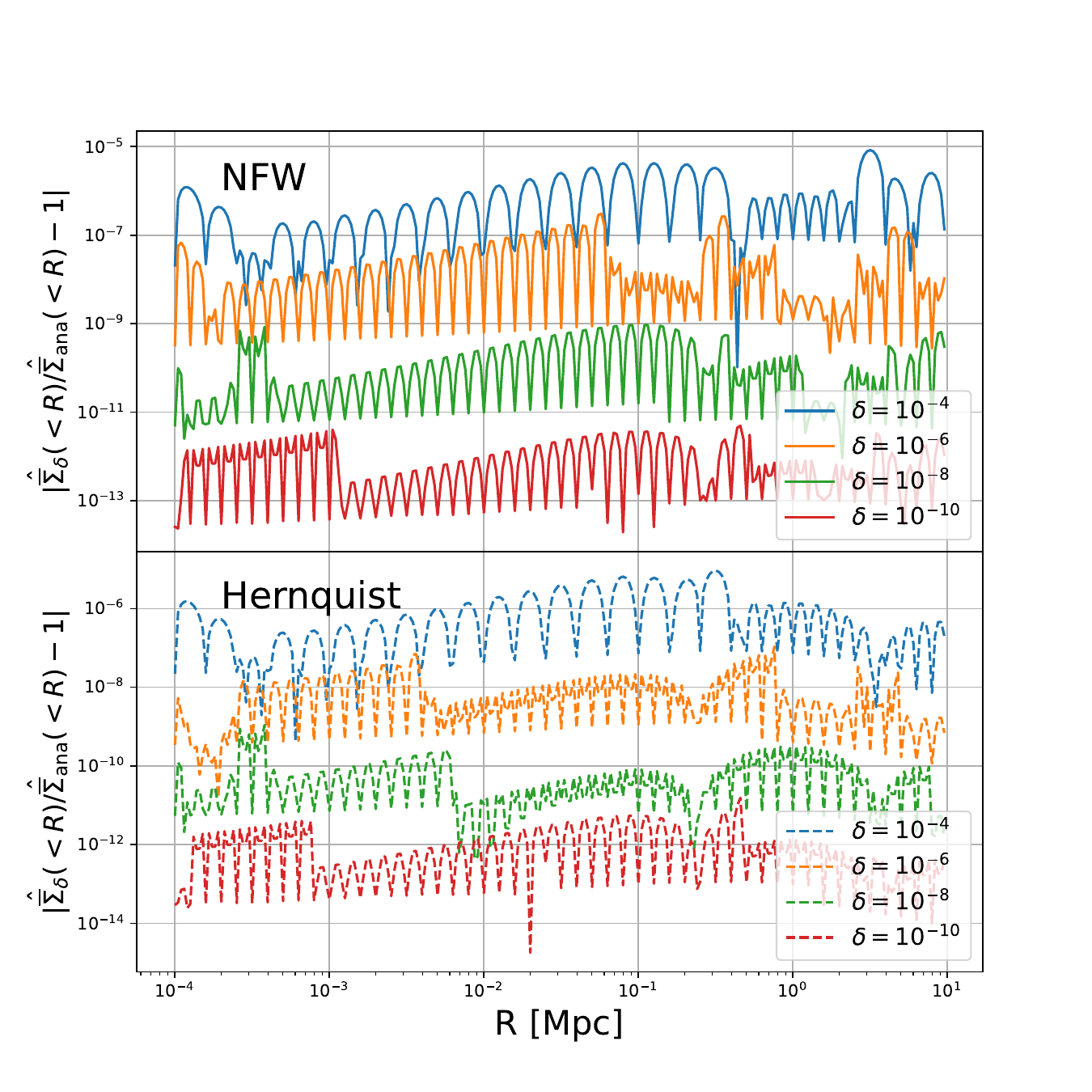}
	\caption{The relative difference between the dimensionless average surface mass
		density within a cylinder of radius $R$, $\hat{\overline{\Sigma}}(< R)$, computed
		using spline interpolation and analytically for the NFW matter density profile
		(upper panel) and the Hernquist profile (lower panel). In both cases, the splines
		were built requiring a relative tolerance $\delta$ equal to $10^{-4}$, $10^{-6}$,
		$10^{-8}$ and $10^{-10}$ corresponding to the blue, orange, green and red curves,
		respectively.}
	\label{fig:MeanSigma}
\end{figure}

In figures~\ref{fig:Sigma} and \ref{fig:MeanSigma}, we present the relative difference
between the analytical and spline-interpolated results for $\hat{\Sigma}(R)$ and
$\hat{\overline{\Sigma}}(<R)$, respectively, for both NFW (upper panel) and Hernquist
profiles (lower panel). In all cases, the interpolated functions achieve the required
precision, typically exceeding one order of magnitude over the entire domain of $R$.

In practice, many cosmological functions do not have analytical solutions, even for
other halo matter density profiles. Consequently, these computations rely on numerical
methods such as integration and ordinary differential equation solvers, often combined
with interpolation techniques. While many codes offer algorithms with hard-coded
interpolation features, such as pre-defined knots, these are typically tested in
specific scenarios. It is impractical to rely on these fixed algorithms to ensure the
necessary numerical precision for statistical analyses, such as Markov Chain Monte Carlo
(MCMC) sampling. The approach presented here addresses this limitation. For example, at
each step of an MCMC run, the functions will meet the required numerical precision by
dynamically generating the list of knots and constructing the corresponding interpolated
functions.

\section{Conclusions and Final Remarks}
\label{sec:conclusion}

This work presents the AutoKnots algorithm, a novel adaptive method for automatic knot
allocation in spline interpolation, designed to satisfy user-defined precision requirements.
In contrast to traditional methods that require manually configured knot distributions
with numerous parameters, the proposed approach automatically determines the optimal
number and placement of knots based on interpolation error criteria.

In section~\ref{sec:tests}, a series of numerical tests were conducted on the AutoKnots algorithm,
demonstrating its consistent ability to accurately
approximate the functions under analysis. While instances of premature convergence were
observed in more complex functions, such as $C_{s}(x)$ and $E_{s}(x)$, where the
adaptive method alone could not achieve the desired tolerance, the refinement process
effectively mitigated these issues, significantly enhancing performance. The number of
knots generated by the adaptive method remained relatively stable across different
functions, with a slight increase when refinement was applied.

The \texttt{refine} option, described in section~\ref{sec:refine}, was found to be
particularly useful when higher accuracy is required. Based on numerical tests, the
recommended configuration is \texttt{refine} = 1 and \texttt{refine\_ns} = 1, as these
values introduce additional knots that improve the adaptive method’s performance for
more complex functions. While the impact on smooth functions was minimal, for more
variable functions, it reduced the tails in the distribution of the maximum relative
error, bringing the results closer to the desired tolerance. However, a drawback is that
about 10\% of unnecessary knots may be introduced for functions that have already met the
required precision.

The analysis of the scale parameter, $\varepsilon$, in Section~\ref{sec:test_scale}
reveals that its determination is strongly influenced by the characteristics of the
function, particularly its maximum absolute value. This was demonstrated by applying
$\varepsilon = 1$ to the functions $\ln(x)$ and $I_{d}(x)$ from section~\ref{sec:ln} and
section~\ref{sec:test_refine}, respectively. For $\ln(x)$, the effect on the approximation
was minimal compared to $\varepsilon = 0$, but for $I_{d}(x)$, a significant improvement
was observed, with a 40\% reduction in the number of generated knots while maintaining
nearly identical precision. This behavior is consistent not only with our method but
also with other numerical techniques, including quadrature evaluations.

Interpolation-based methods for function approximation often depend on hard-coded node
lists. In this work, we address this challenge with an automatic approach driven by the
desired numerical precision. We present our adaptive method and algorithm, detailing the
convergence criteria for knot placement and the user-defined parameters that control
precision: relative tolerance, scale parameter, \texttt{refine} option, and
\texttt{refine\_ns} parameter.

We then demonstrate the application of this method in NumCosmo, using it to calculate
the surface mass density (and its average) for NFW and Hernquist density profiles. The
results show that the relative tolerance is typically about one order of magnitude
larger than the desired precision.

As numerical calculations in cosmology and astrophysics become more complex and
computationally expensive, the use of interpolation methods has become widespread. Many
existing codes rely on hard-coded node lists, with tests conducted for only a small set
of examples. However, in statistical analyses performed over multi-dimensional parameter
spaces, there is often no control over the numerical precision of these calculations.
This method offers a solution, ensuring analyses are both optimized and efficient while
meeting the required precision.

\acknowledgments

SDPV acknowledges the support of CNPq of Brazil under grant PQ-II 316734/2021-7.

\bibliographystyle{JHEP}
\bibliography{references}

\appendix
\section{Algorithmic Implementation}
\label{app:cubicspline}

\subsection{Cubic Spline Interpolation}

In this appendix, we provide an overview of cubic spline interpolation, with a focus on
the algorithmic implementation. The goal is to present an optimized solution to the
interpolation problem by reformulating it in a linear algebraic framework.

Let $f(x)$ be a function defined over the interval $[a, b]$, with $n + 1$ knots such
that $a = x_0 < x_1 < \dots < x_n = b$, where $i = 0, 1, \dots, n-1$. For each
consecutive pair of knots $(x_i, x_{i+1})$, a third-order polynomial $P_i(x)$ is
constructed, ensuring that the polynomial satisfies the conditions of continuity for
both the first and second derivatives. The general form of $P_i(x)$ is:
\begin{equation}
	\label{splineinterp}
	\begin{split}
		P_i(x) & = a_i + b_i (x - x_i) + c_i (x - x_i)^2 + d_i (x - x_i)^3, \\
		x      & \in [x_i, x_{i+1}],
	\end{split}
\end{equation}
where $ a_i $, $ b_i $, $ c_i $, and $ d_i $ are the coefficients to be
determined. The overall cubic spline function, which interpolates $ f(x) $ over the
entire interval, is defined piecewise as follows:
\begin{align}
	\label{spline}
	P(x) = \left\{
	\begin{array}{ll}
		P_0(x),     & x \in [x_0, x_1),     \\
		P_1(x),     & x \in [x_1, x_2),     \\
		\vdots      &                       \\
		P_{n-1}(x), & x \in [x_{n-1}, x_n].
	\end{array}
	\right.
\end{align}
This piecewise cubic polynomial is the basis of the cubic spline interpolation method,
ensuring smoothness and continuity at the interior knots while matching the function
values at the boundary knots.

To determine the $4n$ polynomial coefficients, we need to impose a set of conditions to
ensure the uniqueness of the spline. The first set of constraints ensures that the cubic
spline matches the function values at the given data points $\{x_i\}$.
Specifically, for each knot $x_i$, the spline must satisfy:
\begin{align}
	\label{eq:constraint1}
	P_i(x_i) = f(x_i) = a_i.
\end{align}
This gives $n+1$ linear constraints, corresponding to the function values at the knots,
for $i = 0, \dots, n$. In addition to matching the function values, the spline must also
be smooth at the $n-1$ interior knots. Since each interior knot $x_i$ is shared by two
adjacent polynomials $P_i(x)$ and $P_{i+1}(x)$, we require that the spline be continuous
and have continuous first- and second-order derivatives at each interior knot. These
continuity conditions are:
\begin{align}
	\label{eq:constraint21}
	P_i(x_{i+1})   & = P_{i+1}(x_{i+1}),   \\
	\label{eq:constraint22}
	P'_i(x_{i+1})  & = P'_{i+1}(x_{i+1}),  \\
	\label{eq:constraint23}
	P''_i(x_{i+1}) & = P''_{i+1}(x_{i+1}),
\end{align}
for $i = 0, \dots, n-2$, where $P'_i$ and $P''_i$ denote the first and second
derivatives of $P_i(x)$, respectively. These conditions ensure that the spline is
$C^2$-continuous across the interval.

Next, we express these continuity conditions in terms of the polynomial coefficients.
The first condition, eq.~\eqref{eq:constraint21}, gives:
\begin{align}
	a_i + b_i h_i + c_i h_i^2 + d_i h_i^3 = a_{i+1}, \label{eq:constraint31}
\end{align}
where $ h_i = x_{i+1} - x_i $. The second condition, eq.~\eqref{eq:constraint22},
leads to:
\begin{align}
	b_i + 2c_i h_i + 3d_i h_i^2 = b_{i+1}, \label{eq:constraint32}
\end{align}
and the third condition, eq.~\eqref{eq:constraint23}, yields:
\begin{align}
	c_i + 3d_i h_i = c_{i+1}. \label{eq:constraint33}
\end{align}
These equations must be satisfied for $i = 0, \dots, n-2$, while no continuity
constraints are required at the last or first knot.

Thus far, we have established $4n-2$ conditions: $n+1$ from matching the function values
at the knots and $3(n-1)$ from the continuity conditions at the interior knots.  At this
stage, the equations derived apply to any cubic spline interpolation. The two
remaining conditions come from
the choice of endpoint conditions. There are four possible types of endpoint conditions:
first derivative, second derivative, quadratic, and not-a-knot
constraints~\citep{Knott}. The first two require continuous first and second derivatives
of the interpolated function, with results depending on derivative
approximations~\citep{Tsai}. In this work, we apply the not-a-knot
condition~\citep{Boor78, Boor85, Alhberg}, which is expressed as:
\begin{align}
	P_0'''(x_1)         & = P_1'''(x_1),         & \Leftrightarrow             &
	                    & d_0                    & = d_1, \label{eq:knot1}       \\
	P_{n-2}'''(x_{n-1}) & = P_{n-1}'''(x_{n-1}), & \Leftrightarrow             &
	                    & \quad d_{n-2}          & = d_{n-1}, \label{eq:knot2}
\end{align}
where $P_i'''(x_i)$ denotes the third derivative of $P_i(x)$ evaluated at $x_i$. These
conditions ensure that the third derivatives at the start and end of the mesh are equal,
effectively treating the knots at these points as non-knot points. In our implementation,
we required at least six (6) knots to apply the not-a-knot condition.

\subsubsection{Formulating the Linear System for Spline Coefficients}

To compute the cubic spline approximation under the knot-a-knot condition, we must solve
for the coefficients $a_i$, $b_i$, $c_i$, and $d_i$ in the cubic polynomial
representation of each segment. The knot-a-knot condition, explained in the preceding
sections, leads to a system of $4n$ linear constraints. In this section, we reformulate
the problem into a matrix equation, reducing the unknowns to the $c_i$ coefficients,
which are the second derivatives at each knot.

We begin by applying eq.~\eqref{eq:constraint1} to eq.~\eqref{eq:constraint31} to
determine the coefficients $a_i$ and establish additional constraints involving $b_i$,
$c_i$, and $d_i$. Substituting eq.~\eqref{eq:constraint1} into
eq.~\eqref{eq:constraint31}, we write:
\begin{align} \label{eq:fi}
	f_i + b_i h_i + c_i {h_i}^2 + d_i {h_i}^3 & = f_{i+1},
\end{align}
where $f_i = f(x_i)$. This equation holds for $i = 0, \dots, n-1$, ensuring that the
spline matches the function values at the knots, including the boundary condition
$P_{n-1}\left(x_n\right) = f\left(x_n\right)$.

Note that eq.~\eqref{eq:constraint32} depends on the difference between the coefficients
$b_i$ of adjacent polynomials, defined as $\Delta b_i \equiv b_{i+1} - b_i$. Because of
this, we cannot directly use it to solve for $b_i$ and eliminate its dependence in
eq.~\eqref{eq:fi}. Instead, we subtract eq.~\eqref{eq:fi} for the indices $i+1$ and $i$,
resulting in:
\begin{align}
	\label{eq:diffb}
	\Delta b_i      & = \triangle^x f_i - \left(c_{i+1} h_{i+1} - c_i h_i\right) - \left(d_{i+1} {h_{i+1}}^2 - d_i {h_i}^2\right), \\
	\Delta f_i      & \equiv f_{i+1} - f_i,                                                                                        \\
	\triangle^x f_i & \equiv \frac{\Delta f_{i+1}}{h_{i+1}} - \frac{\Delta f_i}{h_i}.
\end{align}
Here, eq.~\eqref{eq:diffb} establishes a connection between the coefficients of
consecutive polynomials. It is valid for $i=0, \dots, n-2$, as
eq.~\eqref{eq:constraint32} is not defined at $i=n-1$. Substituting
eq.~\eqref{eq:constraint32} into eq.~\eqref{eq:diffb} gives:
\begin{align}
	\label{eq:diffc}
	c_{i+1} h_{i+1} + c_i h_i = \triangle^x f_i - \left(d_{i+1}{h_{i+1}}^2 + 2 d_i {h_i}^2\right).
\end{align}
However, eq.~\eqref{eq:diffc} still depends on the coefficients $d_i$ and $d_{i+1}$,
which are unknown. To eliminate these variables, we use eq.~\eqref{eq:constraint33} to
express $d_i$ in terms of $c_i$ and $c_{i+1}$. The challenge is that
eq.~\eqref{eq:constraint33} is not defined at $i=n-1$. To address this, we introduce a
new variable $c_n$ and extend eq.~\eqref{eq:constraint33} with an additional constraint:
\begin{align}
	c_{n-1} + 3d_{n-1} h_{n-1} = c_n.
\end{align}
This extension allows us to express $d_i$ in terms of $c_i$ and $c_{i+1}$ for all $i = 0, \dots, n-1$:
\begin{align}
	\label{eq:di}
	d_i = \frac{1}{3} \frac{c_{i+1} - c_i}{h_i}.
\end{align}
Substituting eq.~\eqref{eq:di} into eq.~\eqref{eq:diffc}, we eliminate $d_i$ and express the system
purely in terms of the $c_i$ coefficients:
\begin{align} \label{eq:final}
	\frac{h_{i+1}}{3} c_{i+2} + \frac{2}{3} (h_i + h_{i+1}) c_{i+1} + \frac{h_i}{3} c_i = \triangle^x f_i,
\end{align}
valid for $i = 0, \dots, n-2$. This completes the formulation of the system of equations
for the $c_i$ coefficients.

This is a good point to summarize the current state of the problem. We have reduced the
interpolation task to solving for the $n+1$ coefficients $c_i$ in the system of linear
equations defined by eq.~\eqref{eq:final}. Once the $c_i$ coefficients are determined,
the remaining coefficients $a_i$, $b_i$, and $d_i$ can be computed using
eqs.~\eqref{eq:constraint1}, \eqref{eq:fi}, and \eqref{eq:di}, respectively. However,
the system is currently underdetermined, as it consists of $n-1$ equations for $n+1$
unknowns. To address this, we introduce additional constraints. The not-a-knot
condition, specified by eqs.~\eqref{eq:knot1} and \eqref{eq:knot2}, provides the two
additional equations required to determine the system fully.

The challenge lies in the fact that the not-a-knot condition (or other endpoint
conditions) often imposes constraints on the coefficients $c_i$, which can render
the system singular. To address this, the system must be reformulated to ensure the
coefficients $c_i$ are uniquely determined. However, depending on how the system is
constructed, the tri-diagonal structure of the matrix—crucial for computational
efficiency—may be lost. To maintain both well-posedness and computational tractability,
the coefficient matrix must be carefully constructed to preserve its tri-diagonal
structure.

This is achieved by initially solving for $c_0$, $c_1$, $c_{n-1}$, and $c_n$. The
not-a-knot conditions given in eqs.~\eqref{eq:knot1} and \eqref{eq:knot2} are applied to
eq.~\eqref{eq:di}, resulting in the following equations:
\begin{align}
	\label{eq:c0}
	\frac{c_1 - c_0}{h_0}             & = \frac{c_2 - c_1}{h_1},         \\
	\label{eq:cn}
	\frac{c_{n-1} - c_{n-2}}{h_{n-2}} & = \frac{c_n - c_{n-1}}{h_{n-1}}.
\end{align}
Solving for $c_0$ in eq.~\eqref{eq:c0} yields:
\begin{align}
	\label{eqc0}
	c_0 = \frac{(h_0 + h_1)c_1 - h_0 c_2}{h_1}.
\end{align}
Substituting $ c_0 $ into eq.~\eqref{eq:final} for $ i = 0 $ gives an expression for $c_1$:
\begin{align}
	\label{eqc1}
	c_1 = \frac{3 h_1 \triangle^x f_0 + \left(h_0^2 - h_1^2\right) c_2}{(h_0 + h_1)\left(h_0 + 2 h_1\right)}.
\end{align}
Applying this result to eq.~\eqref{eq:final} for $i = 1$ leads to:
\begin{align}
	\label{eq:c2_c3}
	\left(\frac{2}{3} (h_1 + h_2) + \frac{\beta h_1}{3}\right) c_2 + \frac{h_2}{3} c_3 = -\alpha \triangle^x f_0 + \triangle^x f_1,
\end{align}
where:
\begin{align}
	\label{eq:beta}
	\beta  & \equiv \frac{h_0 - h_1}{h_0 + 2 h_1},                     \\
	\label{eq:alpha}
	\alpha & \equiv \frac{h_1^2}{(h_0 + h_1)\left(h_0 + 2 h_1\right)}.
\end{align}
Similarly, for $i = n-2$, we solve for $c_n$ and $c_{n-1}$:
\begin{align}
	\label{eqcn}
	c_n     & = \frac{(h_{n-2} + h_{n-1})c_{n-1} - h_{n-1}c_{n-2}}{h_{n-2}},                                                                            \\
	\label{eqcn-1}
	c_{n-1} & = \frac{3 h_{n-1} \triangle^x f_{n-2} + \left(h_{n-1}^2 - h_{n-2}^2\right) c_{n-2}}{(h_{n-2} + h_{n-1})\left(2 h_{n-2} + h_{n-1}\right)}.
\end{align}
For $c_{n-2}$, the following equation is obtained:
\begin{equation}
	\label{eq:cn-2}
	\begin{split}
		 & \left(\frac{2}{3} (h_{n-3} + h_{n-2}) + \frac{\gamma h_{n-2}}{3}\right) c_{n-2} + \frac{h_{n-3}}{3} c_{n-3} = \\
		 & \triangle^x f_{n-3} - \eta \triangle^x f_{n-2},
	\end{split}
\end{equation}
where:
\begin{align}
	\label{eq:gamma}
	\gamma & \equiv \frac{h_{n-1} - h_{n-2}}{h_{n-1} + 2h_{n-2}},                          \\
	\label{eq:eta}
	\eta   & \equiv \frac{h_{n-2}^2}{(h_{n-2} + h_{n-1})\left(2 h_{n-2} + h_{n-1}\right)}.
\end{align}
Finally, with the updated relations for these coefficients, the tri-diagonal matrix
defined by eq.~\eqref{eq:final} for $i = 2, \ldots, n-2$ is represented by $T$:
\begin{equation}\label{eq:coef_matrix}
	T \equiv \begin{bmatrix}
		(2+\beta)h_1 + 2h_2 & h_2           & 0             & \hdots  & \hdots                & 0                             \\
		h_2                 & 2 (h_2 + h_3) & h_3           & 0       & \hdots                & \vdots                        \\
		0                   & h_3           & 2 (h_3 + h_4) & h_4     & 0                     & \vdots                        \\
		\vdots              & 0             & \ddots        & \ddots  & \ddots                & 0                             \\
		0                   & \vdots        & 0             & h_{n-4} & 2 (h_{n-4} + h_{n-3}) & h_{n-3}                       \\
		0                   & 0             & \hdots        & 0       & h_{n-3}               & 2h_{n-3} + (2+\gamma)h_{n-2})
	\end{bmatrix},
\end{equation}
which is a symmetric positive-definite tri-diagonal matrix. The full system of equations
is then given by:
\begin{equation}
	T\times
	\begin{bmatrix}
		c_2     \\
		c_3     \\
		\vdots  \\
		\vdots  \\
		\vdots  \\
		c_{n-3} \\
		c_{n-2}
	\end{bmatrix}
	= 3 \begin{bmatrix}
		-\alpha \triangle^x f_0 + \triangle^x f_1 \\
		\triangle^x f_2                           \\
		\vdots                                    \\
		\vdots                                    \\
		\vdots                                    \\
		\triangle^x f_{n-4}                       \\
		\triangle^x f_{n-3} -\eta \triangle^x f_{n-2}
	\end{bmatrix}.
\end{equation}

This problem is reduced to solving a tri-diagonal system of equations, a computationally
efficient task due to the sparse structure of the coefficient matrix. Since the matrix
is symmetric, we use the LAPACK library~\cite{Angerson1990}, specifically the routine
\texttt{dptsv}, which is optimized for solving symmetric positive-definite tri-diagonal
systems. This routine computes the solution for the coefficients $c_2, \ldots, c_{n-2}$,
ensuring numerical stability and efficiency.

Once these intermediate coefficients are determined, the remaining boundary coefficients
$c_0, c_1, c_{n-1}$, and $c_n$ are computed using eqs.~\eqref{eqc0}, \eqref{eqc1},
\eqref{eqcn}, and \eqref{eqcn-1}. With all coefficients $c_i$ now established, the
polynomial coefficients $a_i$, $b_i$, and $d_i$ for each interval are obtained by
substituting the results into eqs.~\eqref{eq:constraint1}, \eqref{eq:fi}, and
\eqref{eq:di}, respectively. This sequential computation ensures that the cubic spline
interpolation is fully defined, preserving continuity and smoothness across all
sub-intervals while imposing to the not-a-knot boundary conditions.

\subsection{Adaptive Spline Function}
\label{app:ncm_spline_func}

The \texttt{ncm\_spline\_func} algorithm constructs an adaptive spline representation of
a function $f(x)$ by iteratively refining intervals to meet specified error
tolerances. A linked list is employed as the primary data structure, enabling efficient
insertion of new points during the adaptive process. The algorithm starts with an
initial set of points and evaluates the midpoint $\overline{x}_i$ for each interval.
It then checks error criteria based on absolute and integral
estimates~\eqref{eq:convergence}. If the adaptive criteria are not satisfied ($s_i <
	s_\mathrm{conv}$), the interval is subdivided, and $\overline{x}_i$ is inserted into
the list, ensuring that each function evaluation contributes to improving the spline.

When no further refinement is detected, a secondary step is triggered to address
intervals with large segment lengths. This step resets intervals exceeding a threshold
based on the standard deviation of segment lengths for further refinement. By combining
adaptive and statistical criteria, the AutoKnots algorithm efficiently constructs a robust spline
representation that balances accuracy and computational cost, particularly in regions
where higher resolution is required.
\begin{algorithm}
	\caption{ncm\_spline\_func}\label{alg:spline_func}
	\begin{algorithmic}
		\State $\delta \leftarrow [2.22045 \times 10^{-16}, 1]$ \algorithmiccomment{(default: $10^{-8}$)}
		\State $\varepsilon \leftarrow [0, \infty)$ \algorithmiccomment{(default: $0$)}
		\State $s_\mathrm{conv} \leftarrow \mathbb{Z}^+$ \algorithmiccomment{(default: 1)}
		\State \texttt{refine} $\gets \mathbb{Z}^{0+}$ \algorithmiccomment{(default: 1)}
		\State \texttt{refine\_ns} $\gets (0, \infty)$ \algorithmiccomment{(default: 1)}
		\State $t \gets 0$
		\State $\mathbb{K}^t \gets \{ (x_i, f(x_i), s_i = 0) \}_{m=6}$ \algorithmiccomment{\texttt{linked list}}
		\State \texttt{improve} $\gets$ \texttt{True}

		\While{\texttt{improve}}
		\State $\hat{f}^t \gets \text{\texttt{interpolate}}(\mathbb{K}^t)$ \algorithmiccomment{\texttt{cubic-spline-notaknot}}
		\State $\mathbb{K}^{t+1} \gets \mathbb{K}^{t}$
		\State \texttt{improve} $\gets$ \texttt{False}

		\For{$i \gets 0$ \textbf{to} $\text{\texttt{len}}(\mathbb{K}^t) - 2$}
		\State $\overline{x}_i \gets \frac{x_i + x_{i+1}}{2}$ \algorithmiccomment{\texttt{midpoint-rule}}

		\If{$s_i < s_\mathrm{conv}$}
		\State $\bar{f} \gets f(\overline{x}_i)$
		\State $h_i \gets x_{i+1} - x_i$
		\State $\Delta^{\! \mathrm{a}} f \gets \left\vert \bar{f} - \hat{f}^t(\overline{x}_i) \right\vert$
		\State $\hat{\mathcal{I}} \gets \text{\texttt{integral}}(\hat{f}^t, x_i, x_{i+1})$
		\State $\tilde{\mathcal{I}} \gets \text{\texttt{simpson}}(x_i, \overline{x}_i, x_{i+1})$
		\State $\Delta \mathcal{I} \gets \left\vert \hat{\mathcal{I}} - \tilde{\mathcal{I}} \right\vert$
		\If{$\Delta^{\! \mathrm{a}} f \leq \delta \left( \left\vert \bar{f} \right\vert + \varepsilon \right)$ \textbf{and}
			$\Delta \mathcal{I} \leq \delta \left( \left\vert \widetilde{\mathcal{I}} \right\vert + \varepsilon h_i \right)$}
		\State $s_i \gets s_i + 1$
		\Else
		\State \texttt{improve} $\gets$ \texttt{True}
		\EndIf
		\State $\mathbb{K}^{t+1} \gets \mathbb{K}^{t+1} \cup \left\{ (\overline{x}_i, \bar{f}, s_i) \right\}$ \algorithmiccomment{\texttt{insert-after} $i$}
		\EndIf
		\EndFor

		\If{\textbf{not} \texttt{improve} \textbf{and} \texttt{refine} > 0}
		\State $\mathbb{H}^t \gets \{ h_i \}$
		\State \texttt{refine} $\gets$ \texttt{refine} - 1
		\State $\mu \gets \text{\texttt{mean}}(\mathbb{H}^t)$
		\State $\sigma \gets \text{\texttt{std}}(\mathbb{H}^t)$

		\For{$i \gets 0$ \textbf{to} $\text{\texttt{len}}(\mathbb{K}^t) - 2$}
		\If{$h_i > \texttt{refine\_ns} \cdot \sigma$}
		\State $s_i \gets 0$
		\State \texttt{improve} $\gets$ \texttt{True}
		\EndIf
		\EndFor
		\EndIf
		\EndWhile

		\State \textbf{return} $\mathbb{K}^t$
	\end{algorithmic}
\end{algorithm}
\end{document}